\documentclass[useAMS,usenatbib]{mn2e}
\usepackage{epsfig}
\usepackage{graphics}
\usepackage[usenames]{color}
\usepackage{endnotes}

\usepackage{amssymb}
\usepackage{amsmath}
\usepackage{times}
\usepackage{subfigure}
\usepackage{verbatim}
\usepackage{url}

\newcommand{\km}{{\rm km}}
\newcommand{\hz}{{\rm Hz}}

\newcommand{\mpc}{{\textrm{Mpc}}}

\newcommand{\myr}{{\rm Myr}}
\newcommand{\second}{{\rm s}}

\newcommand{\msun}{M_\odot}
\def\lsim{\lower.5ex\hbox{$\; \buildrel < \over \sim \;$}}

\title{Targeting supermassive black hole binaries and gravitational wave sources for the pulsar timing array}

\author[Rosado \& Sesana]{Pablo A. Rosado$^{1}$\thanks{E-mail: pablo.rosado@aei.mpg.de}, Alberto Sesana$^{2}$
\\
$^{1}$ Max Planck Institute for Gravitational Physics, Albert Einstein Institute, Callinstra\ss e 38, 30167, Hanover, Germany \\
$^{2}$ Max Planck Institute for Gravitational Physics, Albert Einstein Institute, Am M\"uhlenberg 1, 14476, Golm, Germany \\
}

\begin{document}

\date{}

\pagerange{\pageref{firstpage}--\pageref{lastpage}} \pubyear{2013}

\maketitle

\title{Targeting supermassive black hole binaries}

\label{firstpage}

\begin{abstract}
This paper presents a technique to search for supermassive black hole binaries (MBHBs) in the Sloan Digital Sky Survey (SDSS).
The search is based on the peculiar properties of merging galaxies that are found in a mock galaxy catalog from the Millennium Simulation.
MBHBs are expected to be the main gravitational wave (GW) sources for pulsar timing arrays (PTAs); however, it is still unclear if the observed GW signal will be produced by a few single MBHBs, or if it will have the properties of a stochastic background.
The goal of this work is to produce a map of the sky in which each galaxy is assigned a probability of having suffered a recent merger, and of hosting a MBHB that could be detected by PTAs.
This constitutes a step forward in the understanding of the expected PTA signal: the skymap can be used to investigate the clustering properties of PTA sources and the spatial distribution of the observable GW signal power; moreover, galaxies with the highest probabilities could be used as inputs in targeted searches for individual GW sources.
We also investigate the distribution of neighboring galaxies around galaxies hosting MBHBs, finding that the most likely detectable PTA sources are located in dense galaxy environments.
Different techniques are used in the search, including Bayesian and Machine Learning algorithms, with consistent outputs.
Our method generates a list of galaxies classified as MBHB hosts, that can be combined with other searches to effectively reduce the number of misclassifications.
The spectral coverage of the SDSS reaches less than a fifth of the sky, and the catalog becomes severely incomplete at large redshifts; however, this technique can be applied in the future to larger catalogs to obtain complete, observationally-based information of the expected GW signal detectable by PTAs.
\end{abstract}


\hyphenation{MBH}
\hyphenation{MBHs}
\hyphenation{MBHB}
\hyphenation{MBHBs}
\section{Introduction}
\label{sec:Introduction}
Studies of the dynamics of nearby galaxies \citep{KormendyRichstone1995,RichstoneEtAl1998} suggest that a supermassive black hole (MBH) must reside at their centers, and there now exists plenty of observational evidence that almost all massive galaxies host a MBH, our Milky Way being the most striking example \citep{GhezEtAl2008,GillessenEtAl2009}.
There is also a variety of investigations that confirm that the mass of the MBH is highly correlated to the mass and velocity dispersion of the bulge of the hosting galaxy \citep{MagorrianEtAl1998,FerrareseMerritt2000,GebhardtEtAl2000,MarconiHunt2003,HaeringRix2004,GrahamEtAl2011,McConnellMa2013}.
In the context of the $\Lambda$ cold dark matter ($\Lambda$CDM) cosmology, large dark matter structures in the universe build up hierarchically \citep{WhiteRees1978}. 
Galaxies form as gas cools at the centers of dark matter halos; small dark matter halos fall onto greater ones, and the galaxies of the former become satellites of the new host. 
At some later time, the smaller galaxies can merge onto the more massive ones, that lie at the bottom of the potential well.
Within this framework, following galaxy mergers, a large number of supermassive black hole binaries (MBHBs) must form along the cosmic history \citep{BegelmanEtAl1980,VolonteriEtAl2003}.

Depending on the mass ratio (i.e., the mass of the satellite galaxy over the mass of the primary) of the two systems, it is customary to divide galaxy interactions in \textit{minor mergers} and \textit{major mergers}.
In a \textit{minor merger}, the satellite is much lighter than the primary and its material can be disrupted before the two centers merge \citep{GuoEtAl2011b}.
Furthermore, dynamical friction \citep{Chandrasekhar1943}, which is the main mechanism that brings the two MBHs towards each other, can become inefficient (causing the merger to take longer than the age of the universe) if the masses of the two galaxies differ too much.
Alternatively, in a \textit{major merger}, the two galaxies have similar masses and their merger can be completed within less than a few Gyr \citep{KitzbichlerWhite2008}.
Once the separation between the two galaxies is smaller than a few tens of kpc, the two MBHs can efficiently transfer energy and angular momentum to the surrounding stars and gas \citep{Quinlan1996,EscalaEtAl2005,DottiEtAl2007,KhanEtAl2011,PretoEtAl2011}, spiraling towards each other.
At pc separations, they form a bound Keplerian binary; when the binary is tight enough (of the order of 0.01 pc) gravitational wave (GW) emission takes over, leading to coalescence.

Gravitational radiation emission from binary systems \citep{MisnerEtAl1973,Maggiore2008} is predicted by Einstein's theory of General Relativity, however, only indirect proofs of this phenomenon have been achieved so far \citep{WeisbergEtAl2010}\footnote{A similar result based on the double pulsar PSR J0737-3039 \citep{BurgayEtAl2003,LyneEtAl2004} by Kramer et al. is in preparation \citep{Kramer2012}.}.
When GW emission becomes the main mechanism of energy loss of a binary, the inspiral process is well described by General Relativity in its lower order, quadrupolar approximation \citep{Peters1964}.
The period of the orbit decreases with time, while the amplitude of the emitted GWs increases.
At the end of the inspiral phase, when the binary approaches the last stable orbit \citep[see for example Box 25.6 of][]{MisnerEtAl1973}, the coalescence occurs, in which the amplitude of the GWs reaches its maximum \citep{AjithEtAl2011}; after this, the binary enters the ring-down phase, and the GW emission rapidly decays.
Merging MBHBs are the most powerful GW emitters in the universe \citep{SesanaEtAl2004}.

The direct detection of GWs, which is the main goal of several state of the art experiments around the world, including GEO600, LIGO, Virgo and KAGRA \citep{PitkinEtAl2011,LIGO2011,LIGO2009,AccadiaEtAl2012,Somiya2012}, will mark the beginning of the era of GW astronomy \citep{Schutz1999,SathyaprakashSchutz2009}.
Many other experiments have been proposed, like eLISA \citep{Shaddock2008,eLISA2013} and ET \citep{PunturoEtAl2010} to fully exploit this new window to the universe, which will unveil valuable information not only about astrophysics, but also about cosmology and fundamental physics \citep{BabakEtAl2011}.

One of the most fascinating means of detecting GWs directly involve the timing of an ensemble of millisecond pulsars (MSPs, pulsars with spinning periods of $\sim 1\,$ms, \citep{Lorimer2008}), forming a pulsar timing array (PTA).
MSPs are the most regular astronomical objects known, and they play a double role in GW astronomy.
On the one hand, they are potential sources of GWs \citep{Prix2009,AnderssonEtAl2011,Rosado2012}; on the other hand, they can also be used as parts of a galactic-scale GW detector.
GWs perturb the space-time metric between the pulsars and Earth, and these small perturbations affect the times of arrival (TOAs) of the pulses \citep{FosterBacker1990,JenetEtAl2005}.
The differences between the expected and measured TOAs are the timing residuals.
By studying the timing residuals, PTAs aim to detect GWs of frequencies between $\sim 10^{-6}\,\hz$ and $\sim 10^{-9} \,\hz$.
There are three independent PTA collaborations around the globe: the EPTA \citep{FerdmanEtAl2010}, NANOGrav \citep{JenetEtAl2009}, and the PPTA \citep{ManchesterEtAl2013}, which work jointly in the IPTA \citep{HobbsEtAl2010}.

Two main sources are expected to contribute to the GW spectrum in the frequency band of the PTA: MBHBs \citep{RajagopalRomani1995,JaffeBacker2003,WyitheLoeb2003} and cosmic strings \citep{RegimbauEtAl2012b,SanidasEtAl2012}.
In particular, the incoherent superposition of the radiation of the numerous MBHBs in the universe may produce a background of GWs \citep{SesanaEtAl2008,Rosado2011}.
Combining our current theoretical models with observational constraints, the present PTA sensitivity limit lies within the $\sim 95\%$ confidence level of the amplitude of the GW background from MBHBs \citep{Sesana2013}; this means that a detection may occur before the end of the decade.

To date, it is still unknown what kind of signal will be detected by the PTA; it can be dominated by the radiation of a handful of individually resolvable MBHBs \citep{SesanaEtAl2009,MingarelliEtAl2012}, or it can be an incoherent superposition of unresolvable sources, i.e. a stochastic background \citep{Maggiore2000b}.
Very efficient searching algorithms have been developed to detect a Gaussian isotropic GW background \citep{VanHaasterenEtAl2011,VanHaasterenEtAl2013}, but the actual properties of the background (especially its isotropy) are currently under investigation \citep{RaviEtAl2012,MingarelliEtAl2013}.
Alternatively, different data analysis techniques are under development to detect the signature of individually resolvable MBHBs \citep{BabakSesana2012,EllisEtAl2012,PetiteauEtAl2013}.
Within this context, it is therefore meaningful to use available observations to better understand the distribution of MBHBs in the neighboring universe.

The goal of this work is to assign to each galaxy of a catalog a probability of containing a MBHB.
By doing that, we complement our theoretical models of galaxy mergers with information about their spatial distribution.
This can be useful for on-going investigations regarding the anisotropy of the GW background \citep{MingarelliEtAl2013,TaylorGair2013}.
A skymap of potential nearby sources of GWs also provides candidates on which a targeted search for GWs can be applied \citep{BurkeSpolaor2013}, and on which algorithms for single MBHB searches can be tested.
From a theoretical perspective, it is interesting to investigate the environments where MBHBs are formed, whether or not they are more likely to be found in galaxy clusters, and the relation of MBHBs with active galactic nuclei \citep[AGN,][]{LyndenBell1969,KauffmannEtAl2003a}.

In order to find a criterion to identify galaxies that may contain a MBHB, we rely on a simulated galaxy catalog.
It is constructed from the Millennium Simulation \citep[MS,][]{SpringelEtAl2005}, using the galaxy formation models from \cite{GuoEtAl2011b} and the all-sky light-cone produced by \cite{HenriquesEtAl2012}, with the stellar population from \cite{BruzualCharlot2003}.
In this fake universe we find that galaxies that suffered a major merger in the ``recent" past (meaning in less than a few hundreds of Myr, which is the time lapse between snapshots of the simulation) have a distribution in redshifts and masses that does not follow that of non-merging galaxies.
Moreover, they present a distinctive statistical distribution of neighbors in their surroundings (to distances up to a few Mpc).
Therefore, we use the mass, the redshift, and the distribution of neighboring galaxies to characterize the signatures of major mergers.

A galaxy that recently experienced a major merger will be referred to as a \textit{B-galaxy}, since it may contain a MBHB.
Conversely, a galaxy that did \textit{not} merge in less than a few hundred Myr, will be called an \textit{N-galaxy}.
Only a fraction of B-galaxies can contain a MBHB, because the binary lifetime is generally shorter than the time lapse between the MS snapshots, which is used to define B-galaxies\footnote{We assume that once the dynamical friction process has ended the two black holes coalesce in an interval of time shorter than a few hundreds of Myr. This assumption is justified in Section \ref{sec:discussion}.}.
B-galaxies containing a MBHB that could be observed by the PTA (when emitting in some frequency interval accessible to the PTA) will be called \textit{PTA-galaxies}.
Once we are able to identify B-galaxies in the fake catalog, we adapt this catalog to the observational limitations of a real catalog (including, for example, the fact that redshifts are affected by peculiar velocities, and the incompleteness of observations at the low mass/luminosity end).
We then perform a similar search on the adapted catalog, and study how the efficiency of the search is affected by these limitations.
Finally, we perform the same search on a real catalog, namely the Sloan Digital Sky Survey (SDSS) seventh data release \citep[DR7,][]{YorkEtAl2000,AbazajianEtAl2009}, and obtain probabilities for real candidates of B- and PTA-galaxies.

The search for B-galaxies is performed using several algorithms.
The simplest of them (based on Bayesian statistics) takes into account only redshifts and masses to characterize galaxies.
When considering the spatial distribution of neighbors around B-galaxies, the search is performed via a machine learning algorithm (MLA).
The method presented here to search for MBHBs in galaxy catalogs provides an alternative to other recent proposals \citep{TsalmantzaEtAl2011,EracleousEtAl2012,ShenEtAl2013,JuEtAl2013}: its advantage is that it is applicable to all galaxies, independent of their emission properties; its disadvantage is that it is a statistical, indirect method, that can be only used to pick candidates which are {\it more likely} to host a binary.
The outcomes of this exploratory study could be improved in several ways (as discussed in Section \ref{sec:discussion}); among other things, the way we adapt our fake catalog to the SDSS' observational limitations is not optimal, which affects the selection of MBHB candidates.

The outline of the paper is as follows.
In Section \ref{sec:catalogs} we describe the galaxy catalogs (both fake and real) employed in this work; here we also explain the process used to adapt the fake catalog to the observational limitations of the real one.
Section \ref{sec:searches} presents the methods applied to assign galaxies a probability of having suffered a major merger in the recent past, and the probability of containing an observable source of GW in the PTA frequency band.
In Section \ref{sec:results} we show the results of our study of the clustering of B- and PTA-galaxies, as well as the efficiency of the different searches.
We also present a skymap of the SDSS galaxies with the largest probabilities of being a B- or a PTA-galaxy.
In Section \ref{sec:discussion}, the main drawbacks of the searches and possible improvements are discussed.
The main achievements, conclusions and caveats of the paper are summarized in Section \ref{sec:conclusions}.
In Appendix \ref{sec:queries} one can find additional material on how the data used in the paper have been obtained (from the MS and SDSS databases).

\section{Description of the catalogs}
\label{sec:catalogs}

\subsection{Real catalog}
\label{subsec:real_catalog}
The real galaxy catalog is the MPA/JHU value-added galaxy catalog\footnote{Maintained by Jarle Brinchmann at \url{http://www.mpa-garching.mpg.de/SDSS/}.} \citep{KauffmannEtAl2003b,BrinchmannEtAl2004,TremontiEtAl2004}, which is based on the SDSS DR7\footnote{\url{http://www.sdss.org/dr7/}} \citep{YorkEtAl2000,StoughtonEtAl2002,AbazajianEtAl2009}.
In fact, we use the stellar masses in the MPA/JHU updated to DR8 photometry\footnote{We thank Jarle Brinchmann for providing us with the updated stellar mass catalog.} \citep{AiharaEtAl2011,AiharaEtAl2011b}.
The SDSS is the largest and most complete redshift survey to date, covering roughly a quarter of the sky.
It contains photometry and imaging of galactic and extragalactic objects, and spectroscopy for a fraction of them.
The MPA/JHU complements the spectroscopy with the stellar masses of galaxies, which are calculated following the prospects of \cite{KauffmannEtAl2003a} and \cite{SalimEtAl2007}.
The maximum redshift measured in this catalog is $z_{\text{max}}=0.7$, which is the maximum redshift considered in our search.
More precisely, the search will be described and performed (in Section \ref{subsec:bayesian_search}) with a maximum redshift of 0.1.
Then, in Section \ref{subsec:extending}, the search will be extended to $z\le 0.7$.
We set a minimum redshift of $z_{\text{min}}=0.01$, because the SDSS imaging is frequently broken up below this redshift \citep{BlantonEtAl2005}; moreover, distances cannot be calculated from the redshift (since the assumption that galaxies drift with the Hubble flow does not hold), instead, one needs to complement the SDSS with other catalogs of nearby galaxies.
The maximum and minimum stellar masses of galaxies found in the real catalog are $m_{\text{max}}=10^{13}\, \msun$ and $m_{\text{min}}=10^6\, \msun$, respectively.

\begin{figure}
\includegraphics{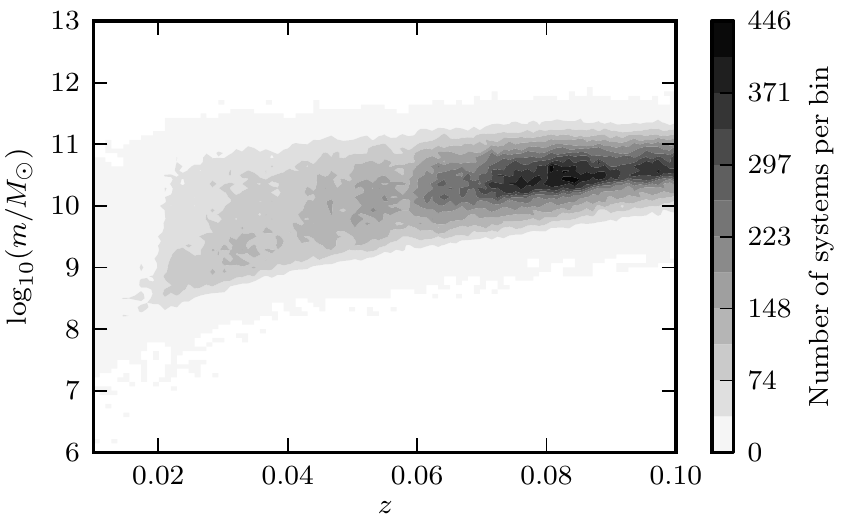}
\caption{Contour plot of stellar mass versus apparent redshift of the real catalog.
The horizontal axis is divided into 100 redshift intervals (or $z$-bins), and the vertical one into 100 intervals of logarithmic mass (let us call them $m$-bins for simplicity).
The gray scale gives the number of systems contained in each redshift-mass pixel (or $z$-$m$-bin).}
\label{fig:zmhist_sdss}
\end{figure}

The SDSS spectroscopic redshift catalog has the advantage of permitting very precise calculations of the positions of galaxies (unlike photometric redshift catalogs, in which redshifts have much larger uncertainties); however, the surveyed area covers only $19.5 \%$ of the sky, and its completeness is affected by several effects \citep{BlantonEtAl2005,GuoEtAl2012}.
Spectroscopic targets of the SDSS are assigned fibers using a tiling algorithm that optimizes completeness \citep{BlantonEtAl2003b}.
But fibers have a finite size, so they cannot be placed close enough to each other in order to aim at targets that have too small angular separations.
Thus, in areas of the sky with high galaxy density, the completeness of the spectroscopic catalog decreases considerably.
This issue, called \textit{fiber collision}, affects the measurements on galaxy clustering \citep{GuoEtAl2012}, specially at scales $\lesssim 1\,\mpc$.
Besides the fiber collision, the completeness of the spectroscopic catalog changes from one region of the sky to another.
All these effects should be taken into account when adapting the simulated catalog to the limitations of the SDSS spectroscopic catalog (see Section \ref{sec:discussion}).
However, we apply a simple method (described in Section \ref{subsec:adapted_catalog}) that does not take them into account.
In Figure \ref{fig:zmhist_sdss} we show how galaxies in the real catalog are distributed in stellar mass and redshift.
This distribution will be used to adapt our fake catalog to the observational limitations of the SDSS (see Section \ref{subsec:adapted_catalog}).

\subsection{Fake catalog}
\label{subsec:fake_catalog}
The MS \citep{SpringelEtAl2005} is an N-body simulation in which $10^{10}$ particles of dark matter evolve in time, in a cubic region of comoving side $\sim 500h_{100}^{-1} \,\mpc$, where $h_{100}=H_0/[100\,\km~\second^{-1}\mpc^{-1}]$ and $H_0$ is the present-day Hubble expansion rate.
These particles interact and form structures in a $\Lambda$CDM universe.
Halos and subhalos are identified using the methods described in \cite{SpringelEtAl2001}, and baryonic matter is then assigned to the halos, following the semi-analytical models of \cite{GuoEtAl2011b}.
The distribution of halos and galaxies is recorded in 64 different snapshots, from redshift $z=127$ to $z=0$.
Since galaxies at $z\gtrsim 0.1$ are distributed in a comoving volume larger than the simulation cube, the latter is repeated periodically.
The mock catalogs (in which equatorial coordinates and apparent redshifts are assigned to galaxies as if we were observers in the fake universe) are constructed as explained in \cite{HenriquesEtAl2012}.

The outcomes of the MS have been contrasted with many observations, confirming that the properties of the fake universe match well the current population of galaxies and MBHs \citep[see for example][]{MarulliEtAl2008,BonoliEtAl2009}.
The cosmological parameters assumed in the MS are a combination of the 2dFGRS \citep{CollessEtAl2001} with the first year of data from WMAP \citep{SpergelEtAl2003}.
To be consistent with the cosmological model of the simulation, when dealing with MS data we assume that the density parameters of matter and dark energy are $\Omega_m=0.25$ and $\Omega_\Lambda=0.75$, respectively, and $h_{100}=0.73$.
Alternatively, SDSS derived data are treated with the cosmological parameters $\Omega_m=0.3$, $\Omega_\Lambda=0.7$, and $h_{100}=0.7$, which are the values assumed in the MPA/JHU.
Neither of these sets of cosmological parameters agree with the most recent measurements; the possible effect of the ``outdated" cosmological parameters in the results is commented on in Section \ref{sec:discussion}.
A new release of the mock galaxy catalogs has been made public during the writing of this paper; the simulation has been rescaled \citep{GuoEtAl2013} to adapt the results to a cosmology based on the data of WMAP 7 \citep{KomatsuEtAl2011}.
Redoing this investigation using the ``updated" fake universe may be considered for a future work.

\begin{figure}
\includegraphics{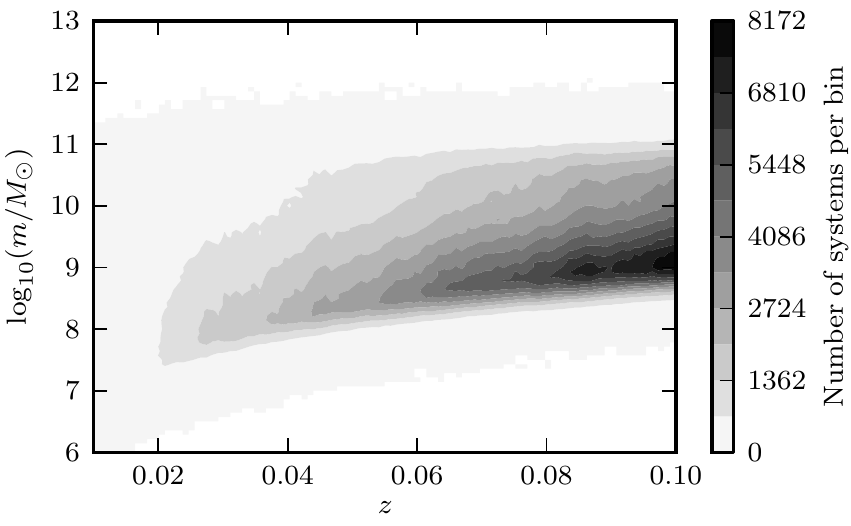}
\includegraphics{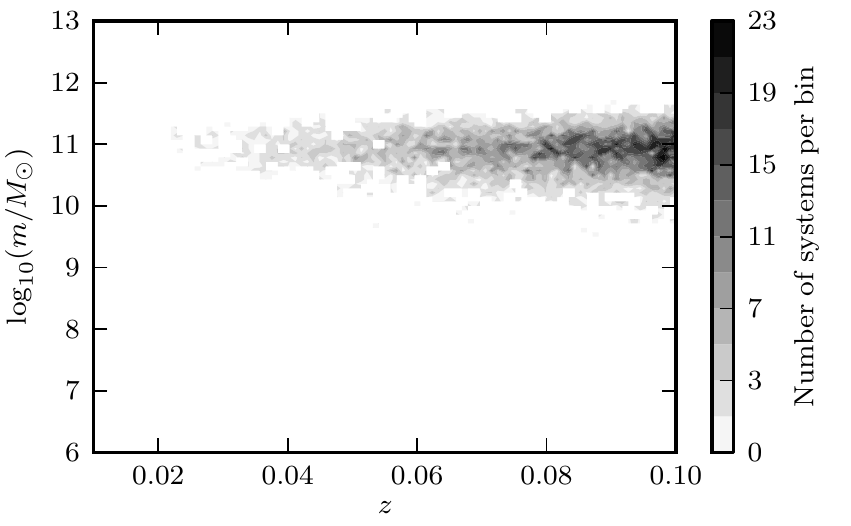}
\caption{Contour plot of stellar mass versus redshift of the fake galaxy catalog (for systems with $z<0.1$).
Both axes are divided into 100 equal bins; the gray scale gives the number of systems contained in each $z$-$m$-bin.
The upper plot considers all galaxies in the fake catalog, while the lower plot contains only B-galaxies.}
\label{fig:zmhist}
\end{figure}

Our fake galaxy catalog can be downloaded from the MS website\footnote{\url{http://www.mpa-garching.mpg.de/millennium/}} \citep{LemsonEtAl2006}, using the SQL query given in Appendix \ref{sec:queries}.
Each galaxy in the fake catalog has a unique identification number (called \verb|galID|).
However, as already pointed out, the simulated universe has a cubic finite size; this cube is repeated periodically, to permit galaxies at larger distances.
A galaxy in one of the cubes has the same mass (as well as other intrinsic properties) and \verb|galID| of its analogous ones in other cubes, but a different sky position and redshift.

B-galaxies are \textit{descendants} of two (or rarely three) merging \textit{progenitors}.
A descendant suffered a \textit{major merger} if the mass of at least two of the progenitors is $\ge 0.2$ times the mass of the descendant.
Moreover, the merger had to occur between the snapshot corresponding to the redshift of the descendant and the immediately previous snapshot.
In Appendix \ref{sec:queries} one can find more details on the selection of B-galaxies, and the query used to download their \verb|galID|.
Once we know which systems are B-galaxies, we can perform searches for B-galaxies in the fake catalog to check how well the methods work.
These searches (described in Section \ref{sec:searches}) are first applied to the local universe (with a maximum redshift of 0.1); then in Section \ref{subsec:extending} we extend the algorithms to a maximum redshift of 0.7.
For $z<0.1$ we find (using the second query of Appendix \ref{sec:queries}) 8400 B-galaxies, of which $\sim 91\%$ are unique (and the rest are repetitions).

In Figure \ref{fig:zmhist}, redshift-mass histograms are plotted for all galaxies and for only B-galaxies.
There we see that B-galaxies are biased towards larger masses.
The histograms in Figure \ref{fig:zmhist} provide a prescription to distinguish B-galaxies from N-galaxies.
With only this information, one can already assign probabilities of galaxies in the real catalog (assuming that we know their redshifts and masses well enough).
The description of such a search is given in Section \ref{subsec:bayesian_search}.

\subsection{Adapted catalog}
\label{subsec:adapted_catalog}

\begin{figure}
\includegraphics{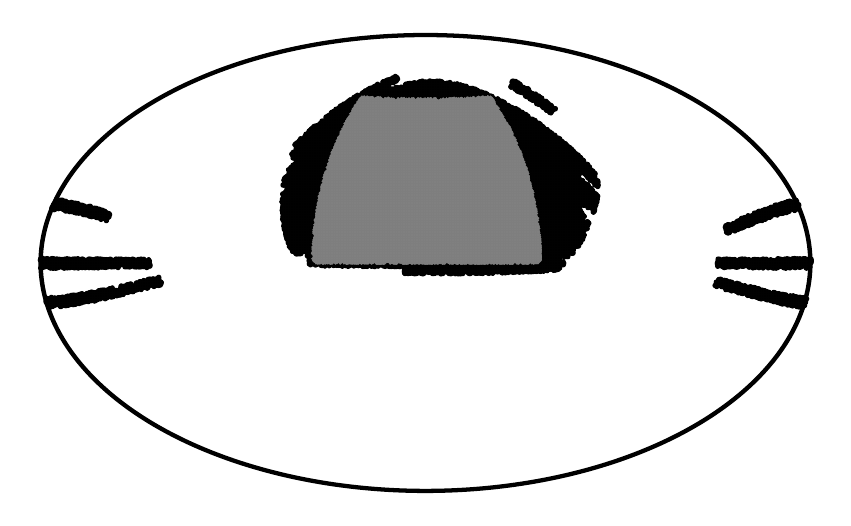}
\caption{Two dimensional (Hammer) projection of the part of the sky covered by our real catalog (black area).
The gray area is the border-free central region we have chosen as model for the redshift and mass distributions of real galaxies, in order to construct the adapted catalog.}
\label{fig:window}
\end{figure}

We now turn to the procedure we have used to adapt our fake catalog to the observational constraints of the SDSS spectroscopic catalog.
As already mentioned in Section \ref{subsec:real_catalog}, this may not be optimal; more sophisticated methods \citep[like the one described in][]{LiEtAl2006b} should be used to properly account for the SDSS incompleteness.

\begin{figure}
\includegraphics{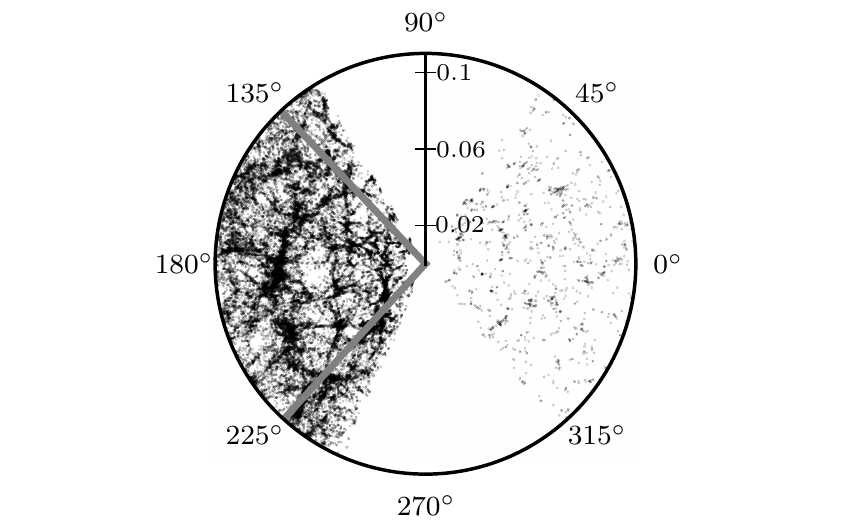}
\includegraphics{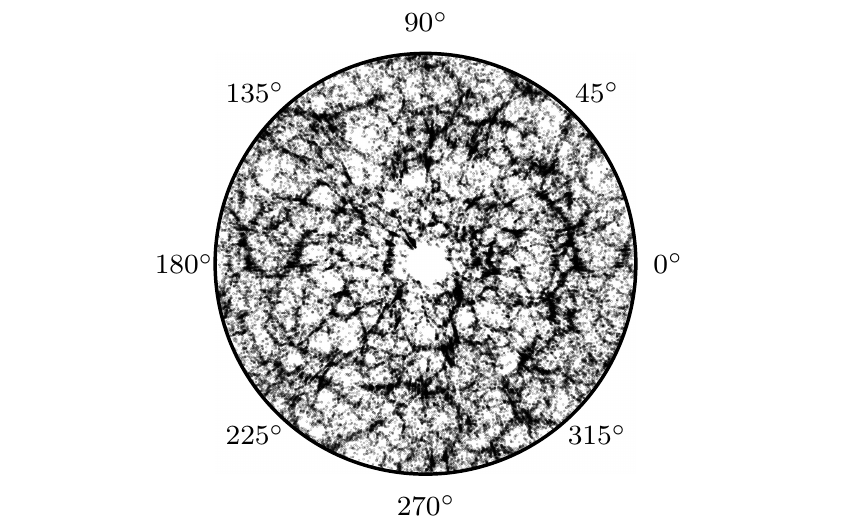}
\caption{Sky distribution of galaxies (projected over the equatorial plane) with right ascension, declination, and redshift in the ranges RA$\in [0^{\circ},360^{\circ})$, DEC$\in[1^{\circ},8^{\circ}]$, and $z\in [0.01,0.11]$, respectively.
The upper plot corresponds to the real catalog, and the lower one to the adapted catalog.
The gray lines in the first plot delimit the region that has been chosen as reference to construct the adapted catalog (the gray region in Figure \ref{fig:window}).}
\label{fig:polarmapSDSS}
\end{figure}

\begin{figure}
\includegraphics{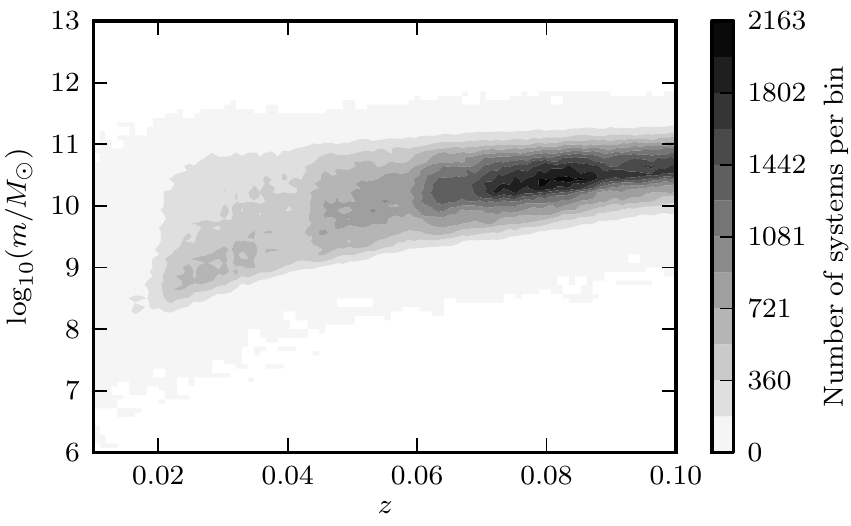}
\includegraphics{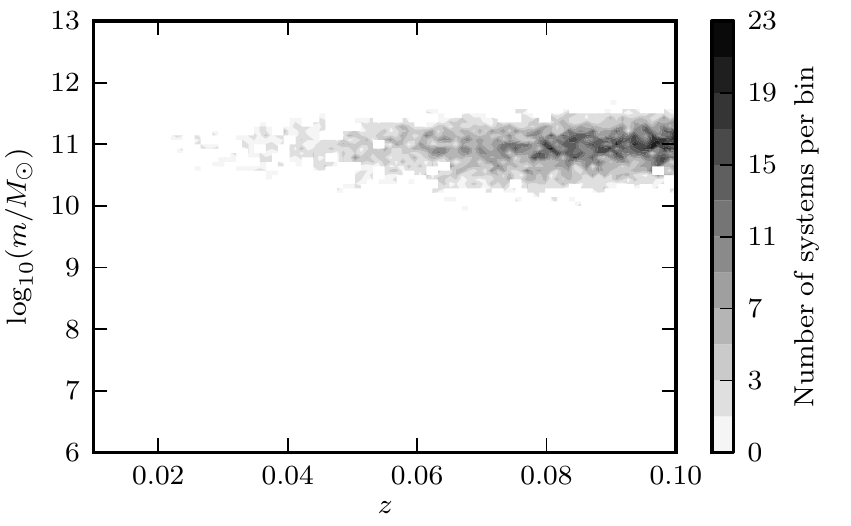}
\caption{Contour plots of stellar mass versus apparent redshift of the adapted catalog.
Both axes are divided into 100 bins; the gray scale gives the number of systems contained in each pixel.
The upper plot considers all galaxies in the adapted catalog.
This plot is to be compared with that of Figure \ref{fig:zmhist_sdss} (but note that the numbers of systems per $z$-$m$ bin are larger here than in Figure \ref{fig:zmhist_sdss}, because the area of the sky covered by the real catalog is smaller than that of the adapted catalog).
The lower plot is obtained for the subset of B-galaxies in the adapted catalog.}
\label{fig:zmhist_adapted}
\end{figure}

Redshifts in the adapted catalog are \textit{apparent redshifts}; these are the redshifts that would be measured if we were observers in the simulated universe, taking into account that galaxies have peculiar velocities.
These apparent redshifts are included in the MS database (labeled \verb|z_app| in the queries of Appendix \ref{sec:queries}).

We first select a region of the sky that is ``almost completely included" in the SDSS spectroscopic catalog, i.e. far from the borders of the SDSS footprint.
In Figure \ref{fig:window}, a projected skymap of galaxies with spectroscopy is plotted.
The selected border-free region is highlighted in gray; this area is limited by two fixed values of right ascension (RA) and declination (DEC), selected in such a way that the area is exactly $1/9$th of the entire sky.
The fake catalog is then divided into 9 regions of equal area.
A $z$-$m$-histogram is calculated for the SDSS central region and for the 9 different patches of the simulated sky.
In all histograms, the axes are divided into 100 bins; where redshifts and masses fulfill $z\in [0.01,0.11]$, and $\log_{10}(m/\msun)\in[6,13]$, respectively.
Each histogram is thus a $100\times100$ matrix in which each element gives the number of systems with redshifts and masses in a particular $z$-$m$-bin.
We compare each of the 9 histograms of the fake catalog with the SDSS histogram.
On the one hand, if one $z$-$m$-bin from the fake catalog contains $n$ more galaxies than the same pixel in the SDSS, $n$ galaxies in that bin are randomly chosen and deleted from the fake catalog.
On the other hand, if one pixel in the fake catalog contains fewer galaxies than the analogous SDSS pixel, nothing is done.
This implies that the number of systems in the adapted catalog is slightly smaller than in the SDSS for some regions of the $z$-$m$-histogram; in fact, at the high-mass end, the fake catalog presents a small shortage of systems with respect to the real one (we will comment on this in Section \ref{sec:discussion}).
Finally, the adapted catalog is the result of combining the 9 sky regions of the fake catalog from which systems have been subtracted.

In Figure \ref{fig:polarmapSDSS}, a map of the real local universe (from the real catalog) is compared to a map of the simulated local universe (from the adapted catalog).
The $z$-$m$-histogram of the adapted catalog is shown in the upper plot of Figure \ref{fig:zmhist_adapted}.
This is, as expected, very similar to that of Figure \ref{fig:zmhist_sdss}, except for the fact that the real catalog contains fewer systems at each pixel, due to the smaller sky region that it covers.
The lower plot of Figure \ref{fig:zmhist_adapted} shows the $z$-$m$-histogram of B-galaxies in the adapted catalog.

\section{Description of the searches}
\label{sec:searches}

\subsection{Probabilities of B-galaxies}
\label{subsec:bayesian_search}
In our searches, each system is characterized by a vector of parameters $\vec{\theta}$.
We start here with the identification of B-galaxies through their peculiar mass and redshift distribution, therefore, for the time being, $\vec{\theta}=\{z,m \}$.
For practical purposes, we divide each parameter range in 100 bins, so that the $z-m$ parameter space forms a matrix of $10^4$ elements.
We name a generic element of this matrix $\theta_i$, where $i=1,...,10^4$.
We now define two functions: $n_f^\text{G}(\theta_i)$ is the number of galaxies (B- or N-galaxies) in the fake catalog with parameters within $\theta_i$; $n_f^\text{B}(\theta_i)$ is the number of B-galaxies in the fake catalog with parameters within $\theta_i$.
The total number of galaxies in the fake catalog is thus
\begin{equation}
\mathcal{N}_f^\text{G}=\sum_i n_f^\text{G}(\theta_i).
\end{equation}
Similarly,
\begin{equation}
\mathcal{N}_f^\text{B}=\sum_i n_f^\text{B}(\theta_i)
\end{equation}
is the total number of B-galaxies in the fake catalog.

The probability of a system in the fake catalog being a B-galaxy, in the case of total ignorance about the parameters $\vec{\theta}$, is
\begin{equation}
\label{eq:prob1}
p_f(\text{B}|I)=\frac{\mathcal{N}_f^\text{B}}{\mathcal{N}_f^\text{G}},
\end{equation}
which is our prior, using the typical notation and nomenclature of Bayesian statistics.
The probability of a system in the fake catalog having $z$ and $m$ within $\theta_i$, given that it is a B-galaxy, is
\begin{equation}
\label{eq:prob2}
p_f(\theta_i|\text{B})=\frac{n_f^\text{B}(\theta_i)}{\mathcal{N}_f^\text{B}},
\end{equation}
which is the likelihood.
Equations (\ref{eq:prob1}) and (\ref{eq:prob2}) can be adapted to the case of N-galaxies; then,
\begin{equation}
p_f(\text{N}|I)=\frac{\mathcal{N}_f^\text{G}-\mathcal{N}_f^\text{B}}{\mathcal{N}_f^\text{G}}
\end{equation}
is the probability of a system being an N-galaxy, in the absence of other information, and
\begin{equation}
p_f(\theta_i|\text{N})=\frac{n_f^\text{G}(\theta_i)-n_f^\text{B}(\theta_i)}{\mathcal{N}_f^\text{G}-\mathcal{N}_f^\text{B}}
\end{equation}
is the probability of an N-galaxy having parameters within $\theta_i$.
Using Bayes' theorem, the probability of a system in the fake catalog being a B-galaxy, given that it has $z$ and $m$ within $\theta_i$, is
\begin{equation}
\label{eq:bayes}
p_f(\text{B}|\theta_i)=p_f(\theta_i|\text{B})\frac{p_f(\text{B}|I)}{p_f(\theta_i|I)}.
\end{equation}
Here, the term in the denominator is the normalization, given by
\begin{equation}
\label{eq:prob3}
p_f(\theta_i|I)=p_f(\theta_i|\text{B})\,p_f(\text{B}|I)+p_f(\theta_i|\text{N})\,p_f(\text{N}|I)=\frac{n_f^\text{G}(\theta_i)}{\mathcal{N}_f^\text{G}}.
\end{equation}
Introducing Equations (\ref{eq:prob1}), (\ref{eq:prob2}), and (\ref{eq:prob3}) in (\ref{eq:bayes}), we get
\begin{equation}
\label{eq:pfb_filt}
p_f(\text{B}|\theta_i)=\frac{n_f^\text{B}(\theta_i)}{n_f^\text{G}(\theta_i)},
\end{equation}
which is an expected result: the probability of a system within a $z$-$m$ bin being a B-galaxy is just the ratio of the number of B-galaxies over the total number of galaxies in that pixel.

The same statistics can be applied to systems in the adapted catalog.
The probability of a system in the adapted catalog to be a B-galaxy, given that it has $z$ and $m$ within $\theta_i$, is
\begin{equation}
\label{eq:pab_filt}
p_a(\text{B}|\theta_i)=\frac{n_a^\text{B}(\theta_i)}{n_a^\text{G}(\theta_i)},
\end{equation}
where we have introduced $n_a^\text{G}(\theta_i)$, the number of galaxies in the adapted catalog with parameters within $\theta_i$, and $n_a^\text{B}(\theta_i)$, the number of B-galaxies in the adapted catalog with parameters within $\theta_i$.
From now on, we will call $p_x(\text{B}|\theta_i)$ the \textit{B-galaxy probability} of a system of catalog $x$ (where $x$ can be `$f$', `$a$', or `$r$', corresponding to the fake, adapted, or real catalog, respectively).

The number of B-galaxies in the real catalog, $n_r^\text{B}(\theta_i)$, is (of course) unknown, but we do know $n_r^\text{G}(\theta_i)$, the number of galaxies in the real catalog with parameters within $\theta_i$.
The function $n_r^\text{G}(\theta_i)$ should be almost identical to $n_a^\text{G}(\theta_i)$ (by construction of the adapted catalog), except for an overall normalization factor (given that the real catalog does not cover the entire sky).
Then, 
\begin{equation}
\label{eq:prb_filt}
p_r(\text{B}|\theta_i)=p_a(\text{B}|\theta_i)
\end{equation}
is assumed to be the probability of a system in the real catalog being a B-galaxy, given that it has $z$ and $m$ within $\theta_i$.

\begin{figure}
\includegraphics{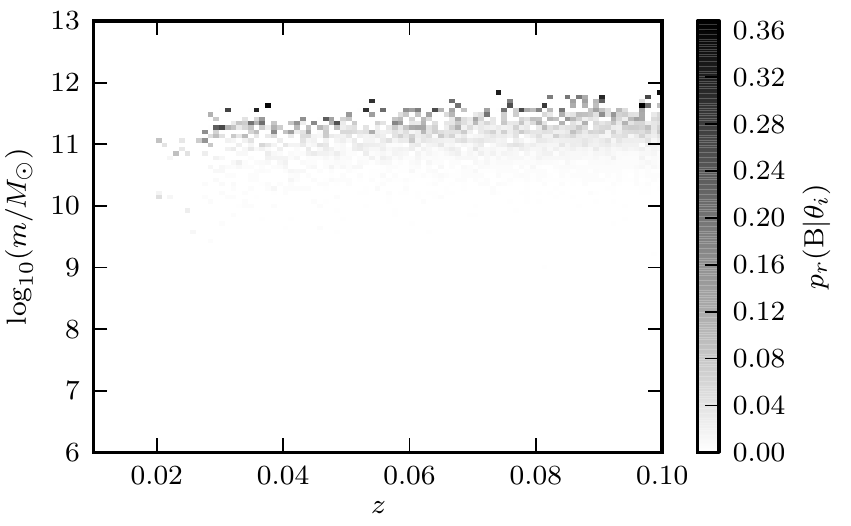}
\includegraphics{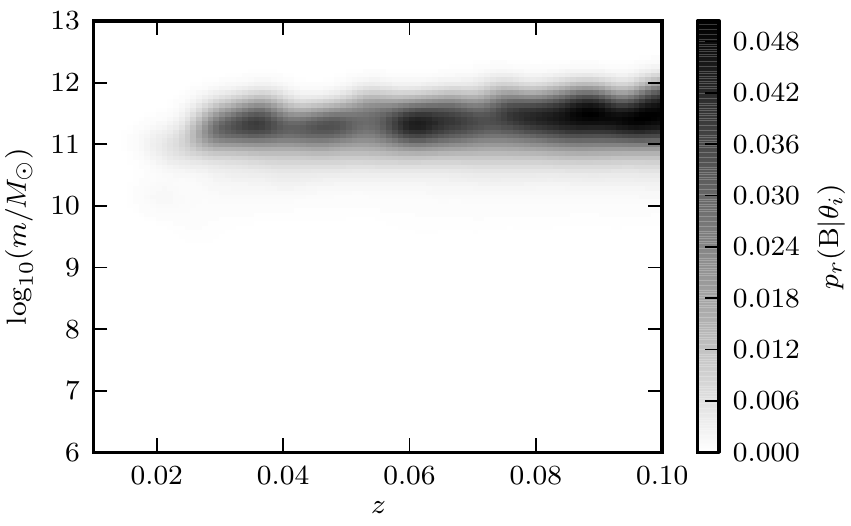}
\caption{B-galaxy probability map as a function of redshift and mass, for systems in the adapted and real catalogs, i.e. $p_r(\text{B}|\theta_i)$.
The upper plot is the (unfiltered) average probability matrix obtained with the galaxies of the training set, as explained in the text.
The lower plot shows the same matrix smoothed with a Gaussian filter; this is the probability matrix that will be used to assign B-galaxy probabilities to systems in the adapted and real catalogs.}
\label{fig:matprob}
\end{figure}

Since we want to test the efficiency of the searches, the probability matrices $p_x(\text{B}|\theta_i)$ (for $x=f$ or $a$) are calculated using systems of only one half of the sky (with $0^{\circ}\le$RA$<180^{\circ}$).
These systems form the \textit{training set}.
Afterwards, the efficiency of the searches are tested (as will be explained in Section \ref{subsec:efficiency}) using systems from the other half (with $180^{\circ}\le$RA$<360^{\circ}$), that form the \textit{testing set}.
Furthermore, the probability matrices $p_x(\text{B}|\theta_i)$ are calculated as the average over 1000 realizations; in each realization, B- and N-galaxies are randomly chosen until covering $19.5\%$ of the sky (the area of the SDSS spectroscopic footprint).
In the case of $p_a(\text{B}|\theta_i)$, in each realization we use a different adapted catalog (from a list of 100 different adapted catalogs, each one built as described in Section \ref{subsec:adapted_catalog}).
This process reduces the amount of systems that are contained both in the training and in the testing sets (because of the repetitions of the simulated cube, mentioned in Section \ref{subsec:fake_catalog}).

An additional remark about the probabilities $p_x(\text{B}|\theta_i)$ needs to be made.
A certain galaxy may have $z$ and $m$ within a bin $\theta_i$ that contains zero B-galaxies, even when all bins around that particular one do contain B-galaxies.
The B-galaxy probability would thus be zero for that system.
But this would not be fair: the probabilities would strongly depend on the sizes of the $z$- and $m$-bins (since, for a different choice of the sizes, that bin $\theta_i$ would not be empty of B-galaxies).
Moreover, our results would also depend too much on the particular realization of the universe that the MS provides.
To avoid these biases, $p_x(\text{B}|\theta_i)$ is smoothed with a two-dimensional Gaussian filter.
The results nevertheless do not change significantly when using different types of filter or using no filter at all.

Finally, each system (from the fake, adapted, and real catalogs) is assigned a value of B-galaxy probability from the smoothed probability matrix $p_x(\text{B}|\theta_i)$, depending on the $z$-$m$-bin $\theta_i$ it falls into.
The unfiltered and filtered probability matrices are plotted in Figure \ref{fig:matprob}.

\subsection{Probabilities of PTA-galaxies}
\label{subsec:search_pta}
A galaxy is said to be a B-galaxy if, among other conditions mentioned in Section \ref{subsec:fake_catalog} or in Appendix \ref{sec:queries}, it is the descendant of two (or more, in the case of a multimerger) progenitors that existed only until the previous snapshot in the simulation.
In other words, between a snapshot and the following one, two (or more) different galaxies became one.
The resulting galaxy may or may not have a MBHB, depending on when the merger actually started and on the lifetime of the binary.
We define the function $T^{\text{snap}}(z)$ as the interval of time elapsed between the current snapshot of a galaxy with redshift $z$ and the previous one (in the fake catalog or in the adapted catalog).

During a galaxy merger several physical mechanisms contribute to shrink the distance between the two MBHs, the final one being the emission of GWs.
At this stage, the time a MBHB spends emitting between two observed GW frequencies $f_1$ and $f_2$ can be calculated using the quadrupolar approximation, and gives
\begin{equation}
\label{eq:timefreq}
\tau(f_1,f_2)=
\frac{5c^5[1+z]^{-8/3}}{256\pi^{8/3}[G\mathcal{M}]^{5/3}}\left[f_1^{-8/3}-f_2^{-8/3}\right].
\end{equation}
Here, $c$ and $G$ are the speed of light and the gravitational constant, respectively, and
\begin{equation}
\mathcal{M}=\frac{[m_\text{BH}^1m_\text{BH}^2]^{3/5}}{[m_\text{BH}^1+m_\text{BH}^2]^{1/5}}
\end{equation}
is the chirp mass of a binary composed of two MBHs of masses $m_{\text{BH}}^1$ and $m_{\text{BH}}^2$.
The lighter the mass of the binary, the longer the time interval it spends emitting in a certain frequency interval.
We disregard the other mechanisms of energy loss that may play an important role when the binary orbits at distances larger than $\sim0.1$ pc.
These mechanisms would enhance the loss of angular momentum at low frequencies, reducing the amount of time the MBHB is emitting GWs \citep{KocsisSesana2011,Sesana2013b}.
The inclusion of these effects in the calculations could be subject of future work.

The GWs produced by a MBHB in a quasicircular orbit, at observed GW frequency $f$, would produce a strain amplitude of
\begin{equation}
\label{eq:strain}
h_0=\frac{2[G\mathcal{M}]^{5/3}\left[\pi f[1+z]\right]^{2/3}}{c^4 r(z)}.
\end{equation}
Here, the function $r(z)$ is the comoving distance between Earth and a galaxy of redshift $z$, given in a $\Lambda$CDM universe by
\begin{equation}
\label{eq:comoving}
r(z)=\frac{c}{H_0} \int_0^z \left[\Omega_m [1+z']^3+\Omega_\Lambda \right]^{-1/2} dz'.
\end{equation}
The values assumed for the cosmological parameters $H_0$, $\Omega_m$, and $\Omega_\Lambda$ are the ones given in Section \ref{subsec:fake_catalog} (we use different parameters for systems from the MS and from the MPA/JHU).
The maximum frequency at which a system can be observed is
\begin{equation}
\label{eq:fmax}
f_{\text{lso}}=\frac{c^3}{6\sqrt{6}\pi G[m_{\text{BH}}^1+m_{\text{BH}}^2][1+z]},
\end{equation}
which is the frequency of the last stable orbit.
The minimum frequency that can be observed is chosen in such a way that the interval of time until the coalescence is not longer than 0.1\,Gyr; we assume that, at lower frequencies, other mechanisms of energy loss would dominate over the GW emission.

The PTA is sensitive to GWs within a certain interval of observed frequencies $[f_\text{min},f_\text{max}]$.
We choose this \textit{frequency window} to be
\begin{equation}
[f_{\text{min}},f_{\text{max}}]=\left[ [10\,\text{yr}]^{-1},[1\,\text{week}]^{-1} \right].
\end{equation}
The lower limit is given by the duration of the PTA campaign; we take 10 years as a default value.
The upper limit is set by the cadence of individual pulsar observations, typically of one per week.
The exact choice of this upper limit does not make any difference in the results, since we do not expect observable sources at such high frequencies anyway.
The PTA frequency window $[f_\text{min},f_\text{max}]$ is divided into 100 frequency bins, equally separated in logarithmic scale.
Galaxies are assigned a probability of being PTA-galaxies at each frequency bin, i.e. they are assigned the probability of producing a strain amplitude larger than a certain threshold within a certain observed interval of frequencies.

Let us first calculate the probability of a system to be a PTA-galaxy at a certain frequency bin, assuming that it is a B-galaxy that contains a binary.
To calculate this probability, $p_x(\text{P}|\text{B};\mathcal{M},f)$, we follow an iterative process.
In each realization (of a total of 100), a total black hole mass $m_\text{BH}$ is drawn from a log-normal distribution, with mean given by the fitting formula from \cite{McConnellMa2013},
\begin{equation}
\label{eq:mbhfun}
\log_{10}\left( \frac{m_\text{BH}}{\msun} \right)=8.46+ 1.05 \log_{10} \left( \frac{m_\text{bulge}}{10^{11}\msun} \right),
\end{equation}
and standard deviation $\sigma=0.34$.
In Equation (\ref{eq:mbhfun}), $m_\text{bulge}$ is the mass of the galactic bulge.
Bulge masses are obtained directly from the MS database for systems in the fake and adapted catalogs (called \verb|bulgemass| in the first query of Appendix \ref{sec:queries}); in Section \ref{subsec:real_galaxies} we explain how to calculate $m_\text{bulge}$ for galaxies in the real catalog.
To avoid considering very light MBH masses\footnote{We point out that the minimum black hole mass found among B-galaxies in the fake catalog is of $10^{6.2}\,\msun$.}, we discard all MBHs with masses smaller than $10^6\,\msun$.
The mass ratio of the binary, $q=m_\text{BH}^1/m_\text{BH}^2$ (with $m_\text{BH}^1\le m_\text{BH}^2$), is drawn at each realization from the distribution of the mass ratios of progenitors' black hole masses in the fake catalog.
Hence, at each realization we have $m_\text{BH}^1=q\,m_\text{BH}$, and $m_\text{BH}^2=m_\text{BH}-m_\text{BH}^1$.
Then, the probability that a galaxy, assuming it contains a MBHB, of catalog $x$ with redshift $z$ is detectable by the PTA for a given minimum strain amplitude $h_0^\text{thres}$ (the strain amplitude threshold) at a certain observed frequency bin $[f_1,f_2]$, is
\begin{equation}
p_x(\text{P}|\text{B};\mathcal{M},f)=\frac{1}{100}\sum_j^{100} p_x^j(\text{P}|\text{B};\mathcal{M},f).
\end{equation}
Here, $f=[f_1+f_2]/2$ is the central frequency of the bin, $j$ denotes the number of the realization, and
\begin{align}
\label{eq:ppzmbh}
&p_x^j(\text{P}|\text{B};\mathcal{M},f)\nonumber\\
&=\left\{ \begin{array}{ll} \min\left(1,\frac{\tau(f_1,f_2)}{T^{\text{snap}}(z)}\right) & \text{if}\quad h_0\ge h_0^\text{thres} \\
0 & \text{if}\quad h_0<h_0^\text{thres}
\end{array} \right. .
\end{align}
In the previous equation we have introduced the function $\min (~)$, to avoid probabilities larger than unity \footnote{This, in practice, does not affect the results, since massive binaries never spend a time exceeding $T^{\text{snap}}(z)$ at an observable PTA frequency interval.}.

Now, applying the product rule, the probability that a galaxy with $z$ and $m$ within $\theta_i$, and MBHB chirp mass $\mathcal{M}$ (if the galaxy hosts a MBHB), is a B-galaxy and a PTA-galaxy in a frequency bin centered at $f$ is
\begin{equation}
\label{eq:pfbp}
p_x(\text{B},\text{P}|\theta_i,\mathcal{M},f)=p_x(\text{B}|\theta_i) \, p_x(\text{P}|\text{B};\mathcal{M},f).
\end{equation}
Finally, the \textit{PTA-galaxy probability} is
\begin{equation}
\label{eq:ptaprob}
p_x(\text{B},\text{P}|\theta_i,\mathcal{M})=\sum_k p_x(\text{B}|\theta_i) \, p_x(\text{P}|\text{B};\mathcal{M},f_k),
\end{equation}
where $f_k$ is the center of a frequency bin; the summation sweeps all frequency bins within the PTA window (as long as the frequency of the last stable orbit, $f_\text{lso}$, is not exceeded, and the interval of time until the coalescence is not longer than 0.1 Gyr).
The PTA-galaxy probability of a system (with $m$ and $z$ within $\theta_i$) in catalog $x$ is, therefore, its probability to be a B-galaxy that contains a MBHB (of chirp mass $\mathcal{M}$) producing a strain amplitude larger than $h_0^\text{thres}$ within the PTA frequency band.

Note that later on we want to apply this machinery to galaxies in the real catalog, where we do not know with certainty which systems are B-galaxies.
For this reason we calculate $p_x(\text{B},\text{P}|\theta_i,\mathcal{M})$ for all galaxies in the different catalogs, even if we know that they are N-galaxies.
Also note that it is meaningless to talk about snapshots of the real catalog (we observe only one snapshot of the universe); however, we expect the probabilities $p_a^j(\text{P}|\text{B};\mathcal{M},f)$ to have a similar behaviour in the simulated universe and in the real one.
Hence we keep the definition of Equation (\ref{eq:ppzmbh}) to calculate $p_r^j(\text{P}|\text{B};\mathcal{M},f)$, where $T^\text{snap}(z)$ is the time between snapshots in the simulation.

\subsection{Including clustering in the search}
\label{subsec:nnds}

We find that B-galaxies tend to cluster differently than N-galaxies (as will be shown in Section \ref{subsec:results_clustering}); this fact motivates us to refine the search by adding information about the clustering of galaxies.
Characterizing galaxies by means of their clustering properties is a common technique in observational astrophysics  \citep[see for example][]{LiEtAl2006a,WangWhite2012}.
These investigations are usually carried out by using the two-point correlation function \citep[TPCF,][]{Peebles1980,Hamilton1993}, which is defined by the joint probability of finding an object simultaneously in two volume elements separated by a certain distance.
The comoving distance between two galaxies is simply calculated as
\begin{equation}
\label{eq:distance}
D_{1,2}=\sqrt{[X_2-X_1]^2+[Y_2-Y_1]^2+[Z_2-Z_1]^2},
\end{equation}
where $(X_j,Y_j,Z_j)$ are the Cartesian coordinates of galaxy $j$ (for $j \in \{1,2\}$), related to the equatorial coordinates by
\begin{equation}
\left\{ \begin{array}{l}
X_j=r(z) \cos(\text{DEC}) \cos(\text{RA})  \\
Y_j=r(z) \cos(\text{DEC}) \sin(\text{RA})  \\
Z_j=r(z) \sin(\text{DEC})   \\
\end{array} \right. .
\end{equation}
These same equations are applied to obtain the positions of galaxies in the fake, the adapted, and the real catalog.
In the adapted and real catalogs $z$ is affected by the peculiar movement of galaxies, but we neglect this effect when calculating distances.
The comoving distance $r(z)$ is defined in Equation (\ref{eq:comoving}), and the cosmological parameters are given in Section \ref{subsec:fake_catalog}.
One can find several definitions for the TPCF in the literature \citep{DavisPeebles1983,LandySzalay1993,Hamilton1993}, which account for possible biases and selection effects of the catalog.
The TPCF is thus a statistical tool that can be used to characterize the clustering of an ensemble of point particles; it is meaningless to talk about the TPCF of an individual galaxy.

Instead, we introduce the \textit{number of neighbors at different shells} (NNDS): this is a set of numbers that measures how many galaxies are contained in spherical comoving shells around a selected object.
For simplicity, we will use the term NNDS both for an ensemble of systems (in which case it denotes the average NNDS over all systems of the ensemble) and for individual systems.
The systems for which the NNDS is calculated (i.e. systems that are at the centers of the shells when counting neighbors) are called \textit{foreground galaxies}.
The rest of the systems (that may be counted as neighbors of some other foreground galaxies) are \textit{background galaxies}.
For the calculation of the NNDS, systems in the {\it adapted} catalog will always be the foreground galaxies; we then investigate two different cases:
\begin{enumerate}
\item All galaxies of the fake catalog are considered as background galaxies.
Redshifts are not affected by peculiar velocities (they are cosmological redshifts).
\item The set of background galaxies is the same as that of foreground galaxies (i.e. galaxies of the adapted catalog).
Redshifts are affected by peculiar velocities (they are apparent redshifts).
\end{enumerate}
The first case corresponds to an idealized case in which we have perfect knowledge of the positions of all galaxies in the universe.
The second is a more realistic approach: the positions of galaxies cannot be precisely calculated (because we do not know the velocity and direction of the peculiar movement of the galaxies), and not all galaxies can be observed.
For both cases, the NNDS is calculated for different \textit{shell sets}, corresponding to different choices of the shells' sizes.
The borders of the shells of some of the sets are separated linearly, and some other logarithmically.
The number of shells in all sets equals 50.

\begin{figure}
\includegraphics{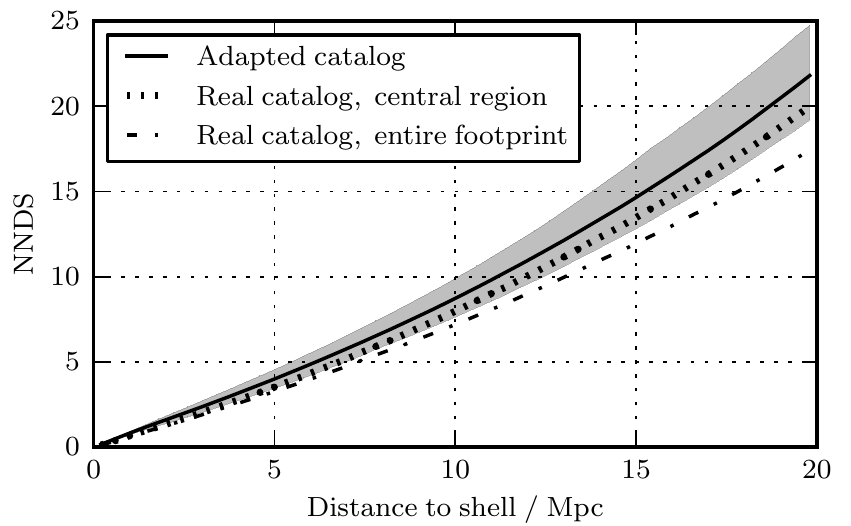}
\includegraphics{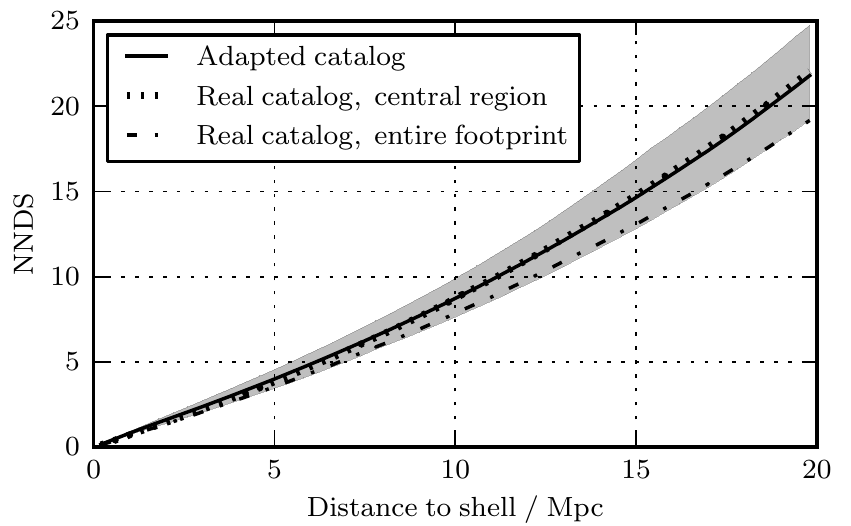}
\caption{Average number of neighbors at different shells (NNDS) of systems in the adapted catalog (solid line), in the real catalog (dot-dashed line), and in a volume of the real catalog that is not affected by the borders of the observed area of the sky (dotted line).
The filled area contains the NNDS of each of the 9 patches in which the adapted catalog is divided.
The number of neighbors is counted for each galaxy at 50 shells with borders equally separated by 400\,kpc.
For the upper plot, distances between systems of the real catalog are calculated assuming the cosmological parameters used by the MPA/JHU, whereas for the lower plot the parameters are those of the MS.}
\label{fig:tpcf_adapted}
\end{figure}

Let us consider case (ii) and a shell set made of 50 shells with borders linearly separated by 400\,kpc.
For each galaxy, we count the number of neighboring galaxies at a distance of less than 400\,kpc; then we count the neighbors that are between 400\,kpc and 800\,kpc away, then between 800\,kpc and 1200\,kpc, etc.
We keep counting neighbors at different shells until reaching a maximum distance of 20\,Mpc (which corresponds to the 50th shell) from the initial galaxy.
Then we average (over all galaxies) the number of neighbors at each shell.
This is shown in the upper plot of Figure \ref{fig:tpcf_adapted}.
The dotted curve shows the NNDS of systems inside the selected central region of the SDSS sky (the gray area in Figure \ref{fig:window}).
The solid line is the NNDS of all systems in the adapted catalog (the filled area contains the NNDS of each of the 9 patches of the sky into which the fake catalog is divided).
The NNDS is also calculated for systems (in the redshift range $[0.01,0.1]$) from the entire real catalog, including the region outside the selected central one (dot-dashed curve).
The latter NNDS is affected by border effects: galaxies close to the border of the observed area of the sky (those galaxies on the black area of Figure \ref{fig:window}) have fewer neighbors.
This effect is increasingly important (and the dot-dashed line differs more from the other two) as more distant shells are considered.
The dotted and solid lines agree quite well, even if there are other sources of incompleteness in the SDSS catalog (besides the effect of the border of the observed sky area) that are not taken into account.

As we mentioned in Section \ref{subsec:fake_catalog}, the MPA/JHU uses different cosmological parameters than the MS.
In the upper plot of Figure \ref{fig:tpcf_adapted}, distances of systems from the real catalog are calculated with the cosmological parameters used by the MPA/JHU; in the lower plot, these distances are calculated with the same parameters used by the MS, and the agreement is much better.
The difference between these plots is the effect of the different sets of cosmological parameters.
In order to properly compare the NNDS of the adapted and real catalogs, it would be convenient to use the updated simulated galaxy catalog of the MS \citep{GuoEtAl2013} (this issue is commented on in Section \ref{sec:discussion}).

In Section \ref{subsec:bayesian_search} we described a search in which each system is characterized by two parameters, $z$ and $m$; we now describe a search, carried out with a MLA, in which galaxies are characterized by a vector of 52 parameters, $\vec{\theta}=\{z,m,\text{NNDS}^1,\dots,\text{NNDS}^{50}\}$, where NNDS$^k$ gives the number of neighbors at the shell $k$ (which goes from 1 to 50).
Machine learning is a growing subject of Artificial Intelligence, and include a vast variety of techniques, in which a program is trained on a number of samples of data (which are characterized by one or more parameters $\vec{\theta}$ called \textit{features}), and tries to predict the characteristics of different sets of data.
We use supervised learning methods of the Scikit-learn\footnote{\url{http://scikit-learn.org}} \citep{PedregosaEtAl2011} library of Python\footnote{\url{http://www.python.org/}} \citep{DrakeVanRossum2011}; in particular, a method that seems to be particularly fast and effective is the Stochastic Gradient Descent.
The algorithm uses a training set as a playground to set up the engine that afterwards assigns a probability of being a B-galaxy, $p_x(\text{B}|\vec{\theta})$, to each of the elements in a testing set\footnote{The mathematical machinery that is used by the algorithm to obtain these probabilities can be consulted at \url{http://scikit-learn.org/stable/modules/sgd.html}.}.

The sky of the simulated universe is divided into 9 patches; the systems in one of them make up our testing set.
We subtract from the rest of the sky all systems which \verb|galID| is contained in the testing set.
This subtraction is performed to avoid systems in the training and in the testing set to be equal (these repetitions would artificially enhance the efficiency of the search).
Then, from the part of the sky outside the testing set, we randomly pick $50\%$ of B-galaxies and $5\%$ of N-galaxies, to construct the training set\footnote{Note that the exact amount of B- and N-galaxies in the training set is not relevant for the results; one just needs to be sure that one has a large enough number of systems of each class to train the MLA properly.}.
The training set is used as input for the MLA.
Then, the MLA calculates the probabilities $p_x(\text{B}|\vec{\theta})$ of systems in the testing set.
In Section \ref{subsec:search_clustering} we show the efficiency of this search.

\subsection{Extending the search to larger redshifts}
\label{subsec:extending}

\begin{figure}
\includegraphics{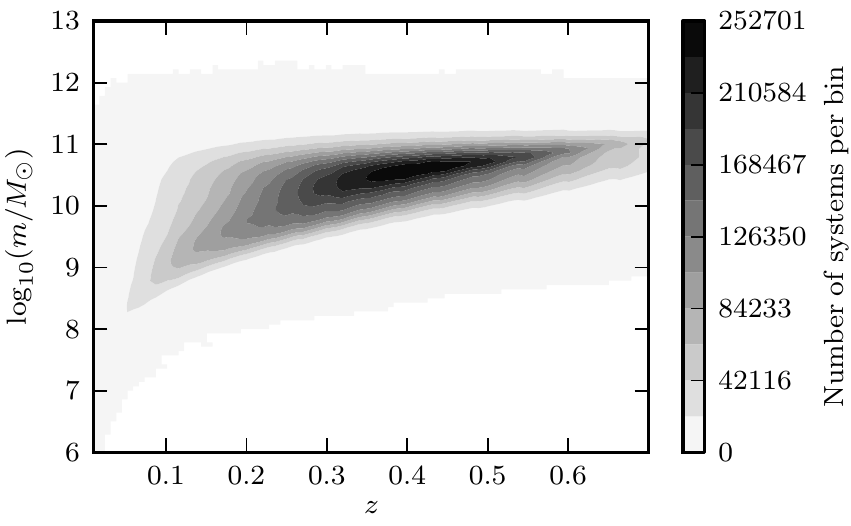}
\includegraphics{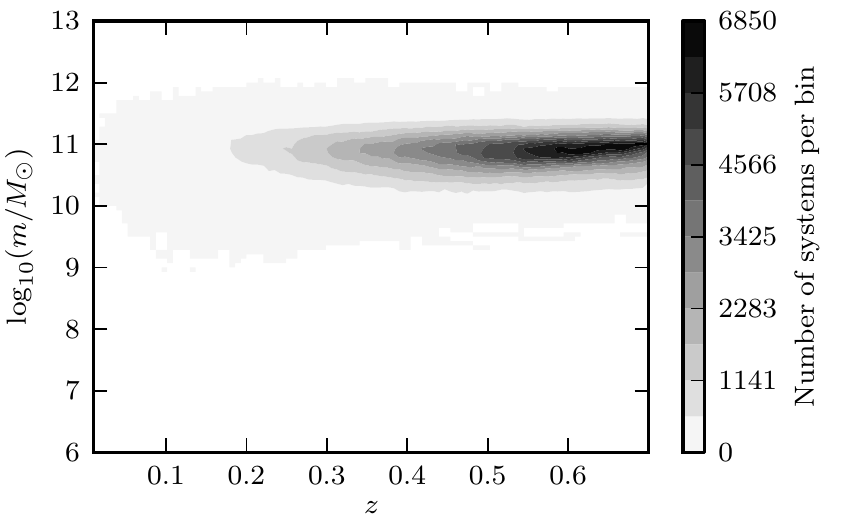}
\caption{Contour plots of stellar mass versus apparent redshift of the entire fake catalog (with $z<0.7$).
Both axes are divided into 100 bins; the gray scale gives the number of systems contained in each pixel.
Upper and lower plots consider all galaxies and B-galaxies, respectively.}
\label{fig:zmhist_larger_z}
\end{figure}

The search in which the clustering is taken into account (described in Section \ref{subsec:nnds}) is performed at redshifts below 0.1.
We do not attempt to extend this search to larger redshifts for several reasons.
Firstly, the completeness of a spectroscopic catalog decreases with distance; the distinct features that are found in the NNDS of B-galaxies will vanish as more neighbors become unobservable.
Secondly, we expect that the method we use to build the adapted catalog will be less trustworthy (regarding the clustering) at larger redshifts.
Finally, as will be shown in Section \ref{subsec:search_clustering}, the inclusion of the NNDS in the search for B-galaxies is already quite ineffective in the local universe; there is no reason why it should improve for $z>0.1$.
Therefore, we spare ourselves the computationally intricate task of calculating the NNDS at larger redshifts.

Conversely, the simple Bayesian search described in Section \ref{subsec:bayesian_search} can be easily extended to $z_\text{max}$.
As Equations (\ref{eq:pfb_filt}), (\ref{eq:pab_filt}), and (\ref{eq:prb_filt}) reveal, the only information we need to assign probabilities is the number of galaxies and B-galaxies within different $z$-$m$-bins.
We do not need to download all systems with $z<0.7$ from the MS database, but just a $z$-$m$-histogram of components $n_f^\text{G}(\theta_i)$, which means $10^4$ integer numbers (for the choice of a $100\times100$ $z$-$m$-grid) with the numbers of galaxies within each pixel, and a histogram of components $n_f^\text{B}(\theta_i)$, with the number of B-galaxies.
In Figure \ref{fig:zmhist_larger_z}, $z$-$m$-histograms of galaxies and B-galaxies from the fake catalog up to $z_\text{max}$ are displayed as contour plots.
For this extended fake catalog, apparent redshifts are used.

\begin{figure}
\includegraphics{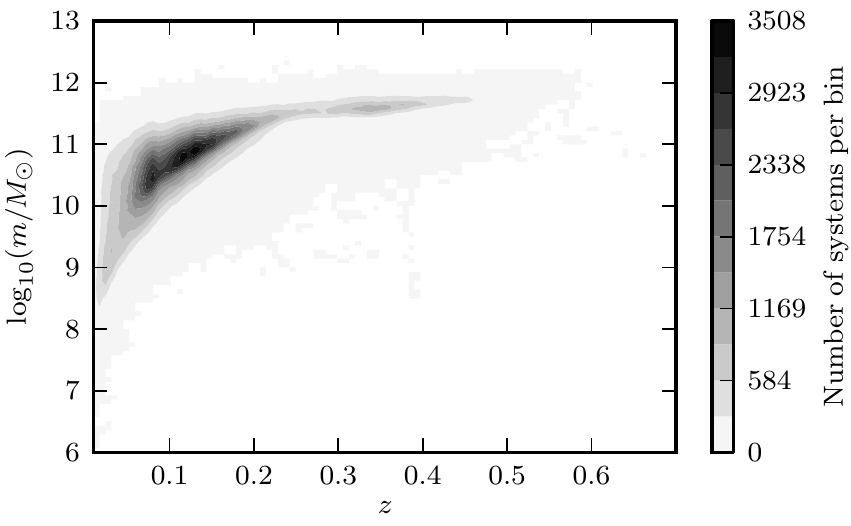}
\includegraphics{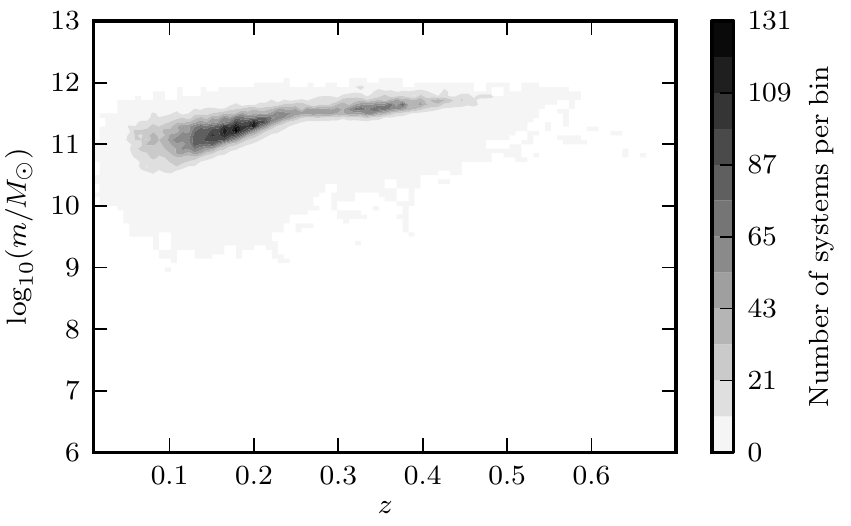}
\caption{Contour plots of stellar mass versus apparent redshift of the extended adapted catalog (that includes systems with $z<0.7$).
Both axes are divided into 100 bins; the gray scale gives the number of systems contained in each pixel.
Upper and lower plots consider all galaxies and B-galaxies, respectively.}
\label{fig:zmhist_larger_z_adapted}
\end{figure}

There is a simple way to construct an adapted catalog using only the functions $n_f^\text{G}(\theta_i)$ and $n_f^\text{B}(\theta_i)$, and the function $n_r^\text{G}(\theta_i)$.
The histogram components $n_a^\text{G}(\theta_i)$ and $n_a^\text{B}(\theta_i)$ can be calculated as
\begin{equation}
n_a^\text{G}(\theta_i)=\min \left(1, \frac{n_r^\text{G}(\theta_i)}{n_f^\text{G}(\theta_i)} \right) n_f^\text{G}(\theta_i),
\end{equation}
and
\begin{equation}
n_a^\text{B}(\theta_i)=\min \left(1, \frac{n_r^\text{G}(\theta_i)}{n_f^\text{G}(\theta_i)} \right) n_f^\text{B}(\theta_i).
\end{equation}
What we are imposing here is that the number of systems in a certain $z$-$m$-bin of the adapted catalog cannot be larger than the same bin of the real catalog.
Figure \ref{fig:zmhist_larger_z_adapted} is analogous to \ref{fig:zmhist_larger_z}, but for galaxies and B-galaxies of the adapted catalog.

Once we have the functions $n_f^\text{B}(\theta_i)$ and $n_f^\text{G}(\theta_i)$ we can construct the B-galaxy probabilities in the extended fake catalog, $p_f(\text{B}|\theta_i)$, using Equation (\ref{eq:pfb_filt}).
Analogously, with the functions $n_a^\text{B}(\theta_i)$ and $n_a^\text{G}(\theta_i)$, the probabilities for the extended adapted and real catalogs can be calculated using Equations (\ref{eq:pab_filt}) and (\ref{eq:prb_filt}), respectively.
Finally, the probabilities $p_x(\text{B}|\theta_i)$ are smoothed with a Gaussian filter, as performed at $z<0.1$.

\section{Results}
\label{sec:results}

\subsection{Clustering of B-galaxies}
\label{subsec:results_clustering}

\begin{figure*}
\includegraphics{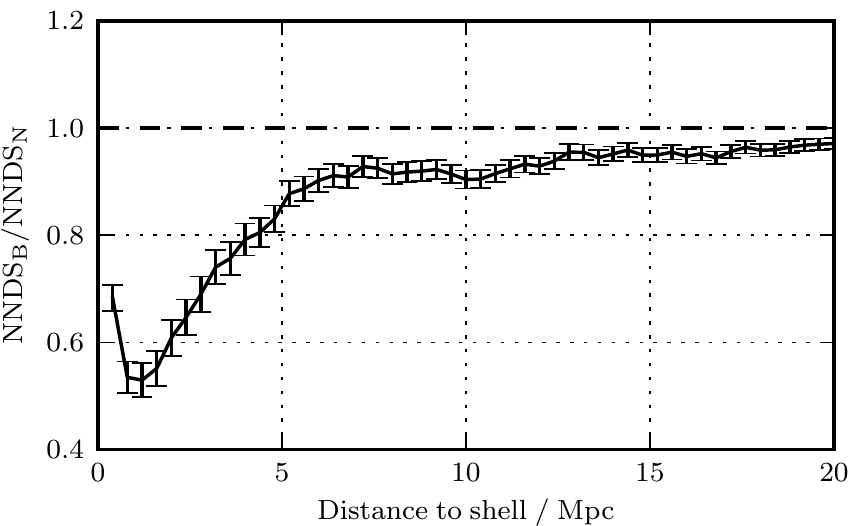}
\includegraphics{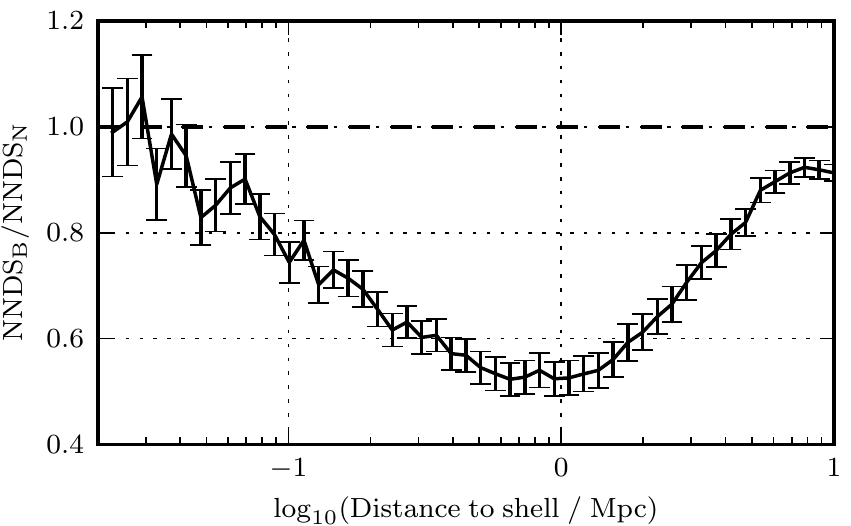}
\includegraphics{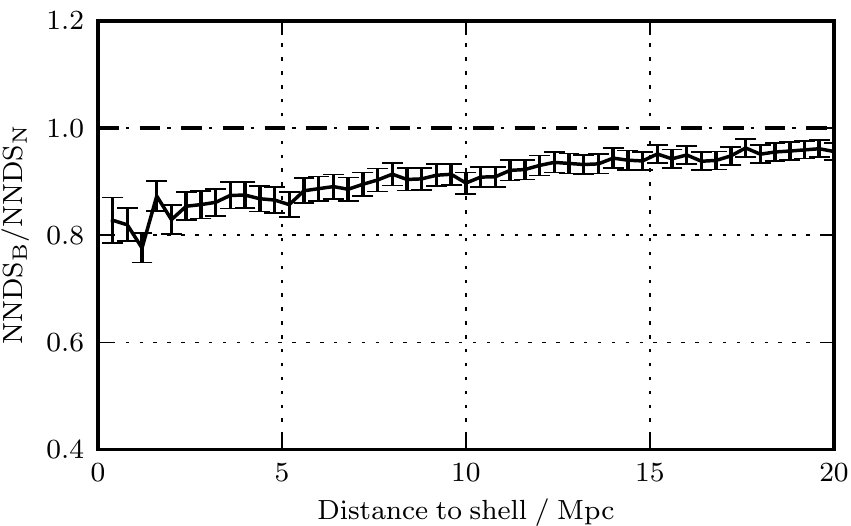}
\includegraphics{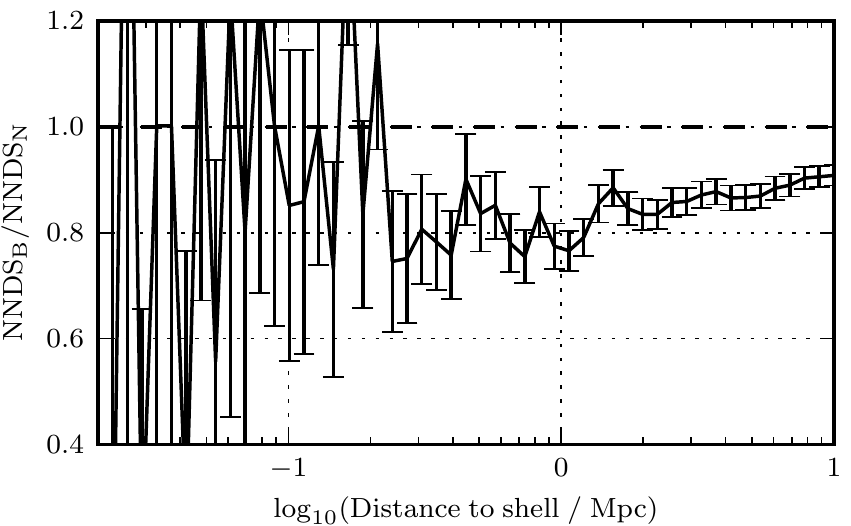}
\caption{Ratio of the NNDS of B-galaxies over the one of N-galaxies, i.e. NNDS$_\text{B}/$NNDS$_\text{N}$, for systems in the fake catalog (upper plots) and in the adapted catalog (lower plots).
N-galaxies are chosen to have the same masses and redshifts as B-galaxies (with matching tolerances of $\Delta \log_{10}(m/\msun)=0.1$ and $\Delta z=0.001$).
Left- and right-hand plots correspond to different shell sets:
shells with borders linearly separated from 0 to 20\,Mpc (left) and logarithmically separated from 1\,kpc to 10\,Mpc (right).
Similar patterns are found when using other shell sets and for other choices of matching N-galaxies.
See the text for an explanation on the sizes of the error bars.}
\label{fig:ratios}
\end{figure*}

\begin{figure}
\includegraphics{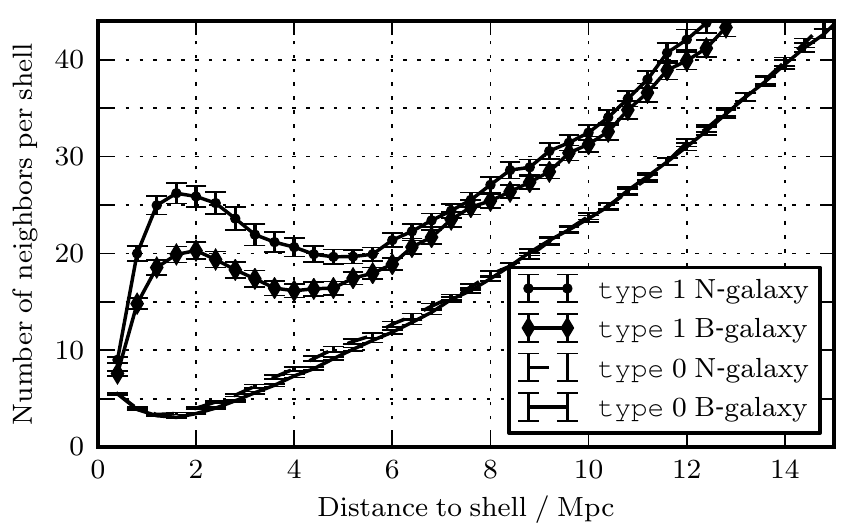}
\includegraphics{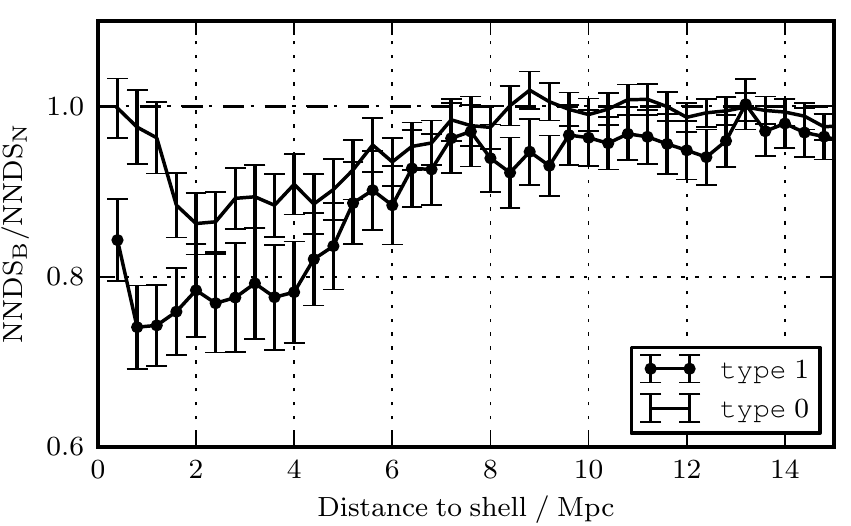}
\caption{The upper plot shows the NNDS of N- and B-galaxies from the fake catalog, of \texttt{type} 0 and 1 (which determines the relative position of the galaxies in their dark matter halos).
Galaxies of \texttt{type} 0 are central galaxies of their halo, whereas galaxies of \texttt{type} 1 are associated with a non-dominant subhalo.
The ratios NNDS$_\text{B}/$NNDS$_\text{N}$ of galaxies of same \texttt{type} are plotted below.
N-galaxies are chosen in such a way that they match redshift, mass, and \texttt{type} of B-galaxies.}
\label{fig:ratios_type}
\end{figure}

\begin{figure}
\includegraphics{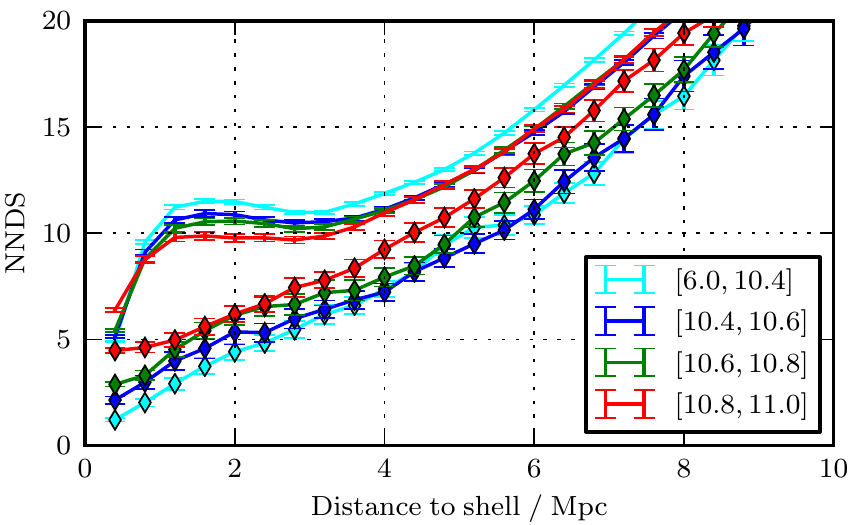}
\includegraphics{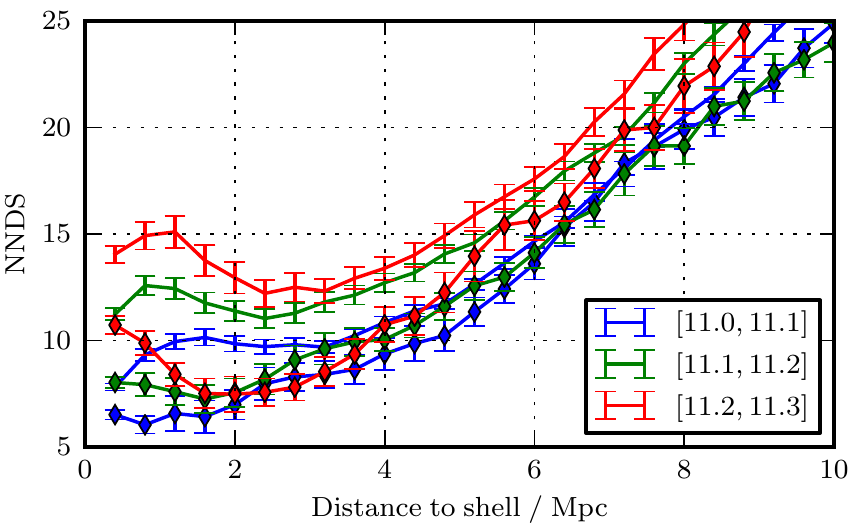}
\includegraphics{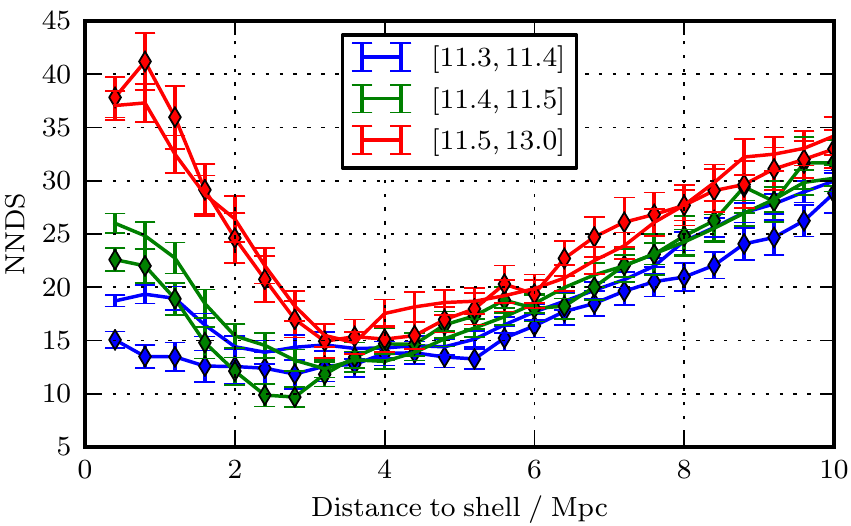}
\caption{NNDS of B- and N-galaxies within different ranges of masses.
Different intervals of masses $m$ (depicted in the legends in units of $\log_{10}\left( m / \msun\right)$) correspond to different colors.
The NNDS of N- and B-galaxies are plotted using points and diamonds, respectively.}
\label{fig:tpcf_mass}
\end{figure}

In Section \ref{subsec:nnds} we mentioned that B-galaxies present a characteristic clustering that could be used to distinguish them from N-galaxies; we now present this distinct shape of the NNDS of B-galaxies, and check how significantly it differs from the NNDS of N-galaxies, when using the fake and the adapted catalogs.
First, the NNDS of B-galaxies, NNDS$_{\text{B}}$, needs to be calculated.
Then, for each B-galaxy, we find an N-galaxy that has similar mass and redshift of that B-galaxy (the matching tolerances are $\Delta \log_{10}(m/\msun)=0.1$ and $\Delta z=0.001$).
The NNDS of these selected N-galaxies is calculated, NNDS$_{\text{N}}$.
The mean NNDS at each shell is obtained both for B- and for N-galaxies.
In Figure \ref{fig:ratios} the ratio NNDS$_\text{B}/$NNDS$_\text{N}$ is plotted; the upper plots correspond to case (i) (as described in Section \ref{subsec:nnds}), in which all systems from the fake catalog can be counted as neighbors; the lower plots correspond to case (ii), and only systems from the adapted catalog can be background galaxies.
The plots on the left are obtained for a set of shells with borders linearly separated by 400\,kpc.
The borders of the shells used for the right plots are logarithmically separated from 1\,kpc to 10\,Mpc.
The error of NNDS$_\text{B}$ and NNDS$_\text{N}$ is assumed to be the square root of the variance of the mean; the error bars in Figure \ref{fig:ratios} are the propagated error of the ratio of both quantities.
We point out that the error bars in the lower right plot are not meaningful at small distances; within those small shells, galaxies in the adapted catalog usually count zero or at most a few neighbors.

There are two reasons why the ratio NNDS$_\text{B}/$NNDS$_\text{N}$ is closer to one for systems in the adapted catalog.
Firstly, distances in the adapted catalog are calculated using apparent redshifts (affected by peculiar velocities), which introduce some level of randomness in the positions of the neighbors.
Secondly, many of the neighbors have been deleted in the process of building the adapted catalog, so the amount of information contained in the NNDS is smaller than when observing all galaxies.

Figure \ref{fig:ratios} shows the interesting fact that B-galaxies present an underdensity of galaxies at $\sim$1\,Mpc scale, when compared to N-galaxies of same redshift and mass.
A similar pattern has already been found in the TPCF of narrow-line AGN in SDSS data \citep[see for example Figure 3 of][but note that, in that paper, projected proper distances are used, instead of spatial comoving distances]{LiEtAl2006b}.
They find that this pattern is also typical of galaxies located at the center of their dark matter halos, where AGN preferentially reside.
It is thus interesting to check whether or not the underdensity of neighbors is due to the relative position of B-galaxies within their halos.
This information can also be extracted from the MS database, by means of the parameter \verb|type| \citep[see Section 3.6 of ][]{GuoEtAl2011b}.
Galaxies of \verb|type| 0 are the principal galaxies of their halos, whereas \verb|type| 1 and 2 are satellite galaxies.
In the fake catalog we find that 80$\%$ of B-galaxies are of \verb|type| 0, and 20$\%$ of \verb|type| 1.
No B-galaxy has \verb|type| 2 (which corresponds to the so-called \textit{orphan galaxies}).
We construct a sample of N-galaxies that match mass, redshift, and \verb|type| of the B-galaxies (this is the reason why the parameter \verb|type| is included in the query of Appendix \ref{sec:queries}), and recalculate NNDS$_\text{B}$, NNDS$_\text{N}$, and the ratio of both.
In the upper plot of Figure \ref{fig:ratios_type} we show the NNDS of B- and N-galaxies of \verb|type| 0 and 1; the ratio NNDS$_\text{B}/$NNDS$_\text{N}$ is shown in the lower plot.
There we see that, in the case of \verb|type| 0 galaxies, the underdensity is due to $\sim$ 1 neighbor at distances of a few Mpc,
hence the clustering of \verb|type| 0 B- and N-galaxies of similar masses and redshifts are almost indistinguishable.
B-galaxies of \verb|type| 1, however, still present a significant underdensity of neighbors with respect to matching N-galaxies at around 1\,Mpc.

In Figure \ref{fig:tpcf_mass} we show the NNDS of B- and N- galaxies at different mass intervals.
Galaxies within the same mass interval have NNDS displayed with the same color; the NNDS of B-galaxies is marked with diamonds.
Here one can clearly see that B-galaxies have on average fewer neighbors (especially at distances of a few Mpc) than N-galaxies of similar masses.
An important feature to note is that the larger the mass is, the larger the number of neighbors becomes, both for B- and for N-galaxies.

An interesting question is whether or not PTA-galaxies are more likely to be found in galaxy clusters.
In a previous section we defined PTA-galaxies as B-galaxies that can contain a MBHB emitting GWs (of frequencies within the PTA window) that produces strain amplitudes larger than a certain threshold $h_0^\text{thres}$.
According to Equation (\ref{eq:strain}), the strain amplitude is proportional to $\mathcal{M}^{5/3}$, so MBHBs need to be very massive to produce large strains.
However, the time a binary spends in the PTA frequency band is proportional to $\mathcal{M}^{-5/3}$, meaning that more massive binaries are less likely to be found in the GW emission phase.
There exists a trade-off between the two arguments, that happens to favor larger masses.
In Section \ref{subsec:real_galaxies}, we will make a list of real PTA-galaxy candidates; their masses lie between $\sim 10^{11.1}\,\msun$ and $10^{12.1}\,\msun$, with a mean of $10^{11.7}\,\msun$.
As Figure \ref{fig:tpcf_mass} shows, galaxies with such masses tend to have significantly more neighbors than average (lower-mass) galaxies.
This argument is not enough to conclude that PTA-galaxies are usually in big galaxy clusters, but we can nonetheless claim that they are more likely located in dense neighborhoods.
A more precise answer to the question of PTA-galaxies being found preferentially or not in clusters could be achieved by performing an exhaustive study of the list of PTA-galaxy candidates of Section \ref{subsec:real_galaxies}.

We remark that it is more customary to use projected distances in studies regarding galaxy clustering and neighbors, rather than three-dimensional distances.
Indeed, the actual magnitude that we are able to observe from two distant objects is their two-dimensional distance (projected on the plane perpendicular to the line of sight), and it is often not possible to measure the redshift of all neighbors of a particular galaxy.
Nevertheless, since the clustering study of this paper is restricted to the simulated universe (in which all spatial information is provided), the use of three-dimensional distances is sufficient.
A more detailed investigation of the clustering properties of B-galaxies, using projected distances, may be subject of a forthcoming work.

\subsection{Efficiency of the search for B-galaxies}
\label{subsec:efficiency}

\begin{figure}
\includegraphics{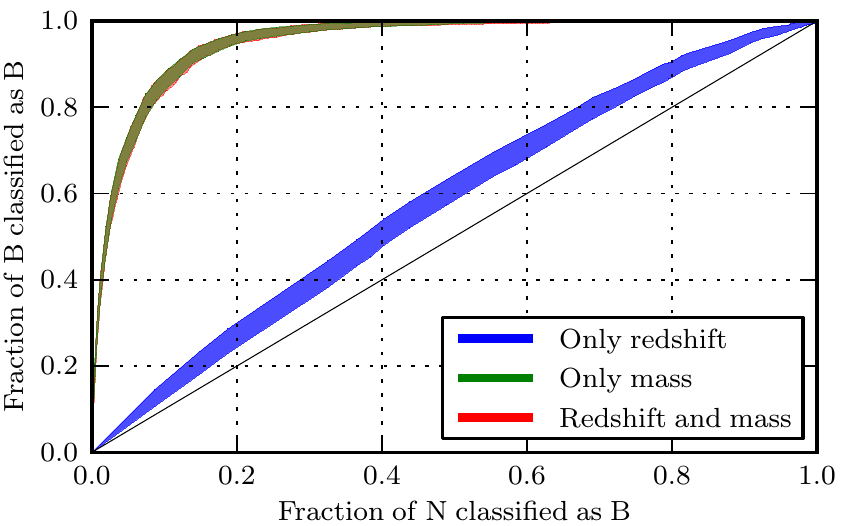}
\caption{Results of 1000 realizations of the search for B-galaxies in the fake catalog using the simple Bayesian approach described in Section \ref{subsec:bayesian_search}.
The colored areas contain the ROC curves produced in each of the 1000 realizations.
The red area contains the curves obtained when both mass and redshift are considered as parameters in the search.
The green (or blue) areas contain the ROC curves obtained when characterizing galaxies only by their mass (or redshift).
The brown area is the overlap of the red and green ones, that are almost identical.}
\label{fig:ROC_MS_bayesian}
\end{figure}

\begin{figure}
\includegraphics{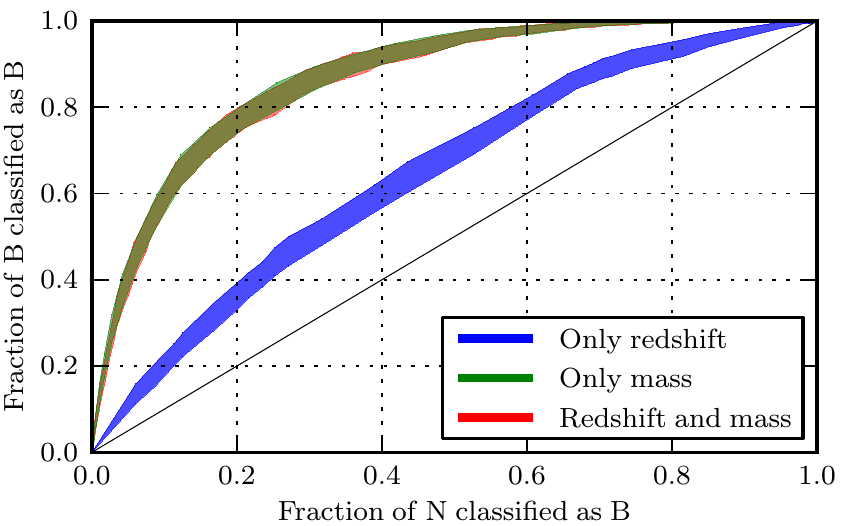}
\caption{Analogous to Figure \ref{fig:ROC_MS_bayesian}; now only systems from the adapted catalog are considered.}
\label{fig:ROC_adapted_bayesian}
\end{figure}

In Section \ref{subsec:bayesian_search} we described how a training set of galaxies (chosen from one half of the simulated sky) is used to create a matrix of probabilities $p_x(\text{B}|\theta_i)$; we now address the question of how well we can identify B-galaxies from the testing set (constructed with systems of the other half of the sky).
Galaxies are randomly chosen from the testing set, until covering $19.5\%$ of the sky.
We define a vector of threshold probabilities $p_T$, with components in the range $[0,1]$.
For each value of $p_T$, we count the number of N-galaxies classified as B-galaxies (i.e. with $p_f(\text{N}|\theta_i)>p_T$) and the number of B-galaxies classified as B-galaxies (with $p_f(\text{B}|\theta_i)>p_T$).
This process is repeated 1000 times.

In Figure \ref{fig:ROC_MS_bayesian} we plot \textit{receiver operating characteristic} (ROC) curves of the search in the fake catalog.
A ROC graph represents the false alarm rate (number of N-galaxies classified as B-galaxies divided by the number of N-galaxies in the testing set) on the horizontal axis and the detection rate (number of B-galaxies classified as B-galaxies divided by the number of B-galaxies in the testing set) on the vertical one.
The thin black line that crosses the plot diagonally would be the result of a totally random search (for each probability value, one gets the same fraction of good and bad classifications).
The red area contains 1000 ROC curves (each one corresponding to a different testing set) for which galaxies are characterized by the parameters $\vec{\theta}=\{z,m\}$.
The green area contains also 1000 ROC curves; in this case galaxies are characterized only by their mass, so $\vec{\theta}=\{m\}$.
Finally, the blue area is filled with 1000 ROC curves for which galaxies have been characterized by their redshift, $\vec{\theta}=\{z\}$.
This plot makes clear that the most useful piece of information to distinguish B-galaxies is their mass.
The same procedure to test the efficiency of the search is applied to systems in the adapted catalog.
Figure \ref{fig:ROC_adapted_bayesian} is analogous to Figure \ref{fig:ROC_MS_bayesian}, but now the training and testing sets are constructed with systems from the adapted catalog.

\begin{figure}
\includegraphics{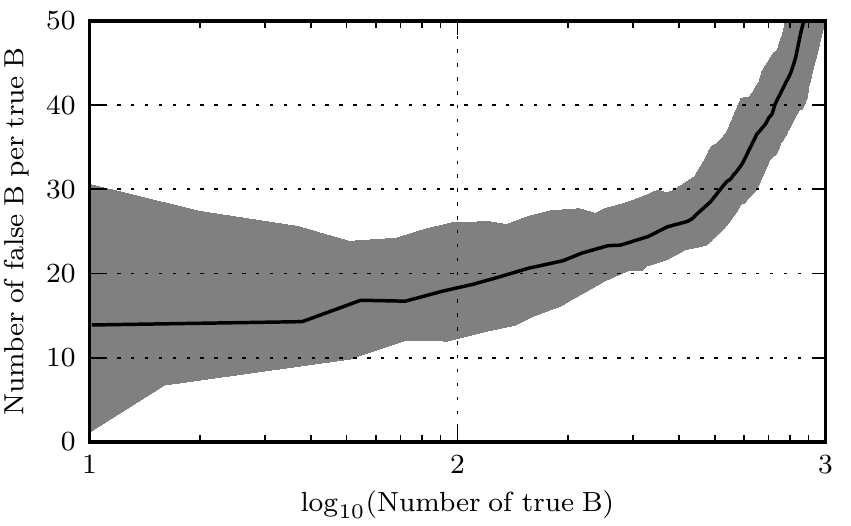}
\caption{Number of wrong classifications per good one versus well-classified B-galaxies from the adapted catalog, after using the search described in Section \ref{subsec:bayesian_search}.
This figure is a different representation of the data contained in Figure \ref{fig:ROC_adapted_bayesian}.
The gray area contains the curves corresponding to the 1000 realizations of the search (each one for a different choice of the galaxies in the testing set).
The black curve is the average over the 1000 curves; each point corresponds to a value of B-galaxy probability.}
\label{fig:threshold_bgal}
\end{figure}

ROC curves are commonly used to test the efficiency of classification algorithms, however, for our purposes, we find more convenient to present the data in a slightly different manner.
In Figure \ref{fig:threshold_bgal} we present the number of well classified B-galaxies from the adapted catalog (on the horizontal axis) and the number of incorrectly classified N-galaxies per well-classified B-galaxy (on the vertical axis).
The gray area contains the curves obtained for the 1000 realizations; the black line is the average over all of them.
Each point of a curve corresponds to a certain value of probability $p_a(\text{B}|\theta_i)$.
The points on the right part of the plot (where more B-galaxies are well classified) correspond to lower probabilities.

A probability threshold value $p_T^\text{B}$ needs to be chosen, so that all galaxies with $p_a(\text{B}|\theta_i)\ge p_T^\text{B}$ are considered \textit{candidates} for B-galaxies.
We choose a threshold such that an average of $\sim100$ B-galaxies are counted as candidates.
More precisely, for $p_T^\text{B}=4.16\times 10^{-2}$, in the worst realizations we find 82 B-galaxies among 2166 candidates, whereas in the best ones, 143 B-galaxies out of 2106 galaxies are found.
The average result produces 110 B-galaxies among 2168 candidates.
Further on we apply this same search to galaxies in the real catalog.
B-galaxy candidates are chosen using the same probability threshold $p_T^\text{B}$, so we expect to have a similar number of galaxies with $p_r(\text{B}|\theta_i)\ge p_T^\text{B}$, and thus a similar number of real B-galaxies among them too.
We point out that our choice of $p_T^\text{B}$ is arbitrary.
One could have chosen a smaller one, and more B-galaxies would be counted among the candidates; nonetheless, as Figure \ref{fig:threshold_bgal} shows, for a smaller threshold the ratio of N-galaxies per B-galaxy considered as candidates would also be larger.

\subsection{Efficiency of the search for PTA-galaxies}
\label{subsec:efficiency_pta}

We now test the efficiency of the search for PTA-galaxies in the adapted catalog, following a similar procedure to the one explained in Section \ref{subsec:efficiency}.
The strain amplitude threshold is set to $h_0^\text{thres}=10^{-15}$; with this we calculate, for each system, its PTA-galaxy probability, $p_x(\text{B,P}|\theta_i,\mathcal{M})$.
Because of the iterative nature of the procedure used to assign PTA-galaxy probabilities, systems cannot be repeated: even if two galaxies have the same \verb|galID| and $m$, the masses of their MBHs will be different, leading to different probabilities.
For this reason the adapted catalog is not divided into a training and a testing set.
Galaxies (that could contain a MBHB producing a strain amplitude larger than $h_0^\text{thres}$) are chosen randomly from the whole simulated sky until covering an area of the sky of $19.5\%$; we then count the number of B- and N-galaxies passing each value of the probability threshold.
This procedure is repeated 1000 times.

\begin{figure}
\includegraphics{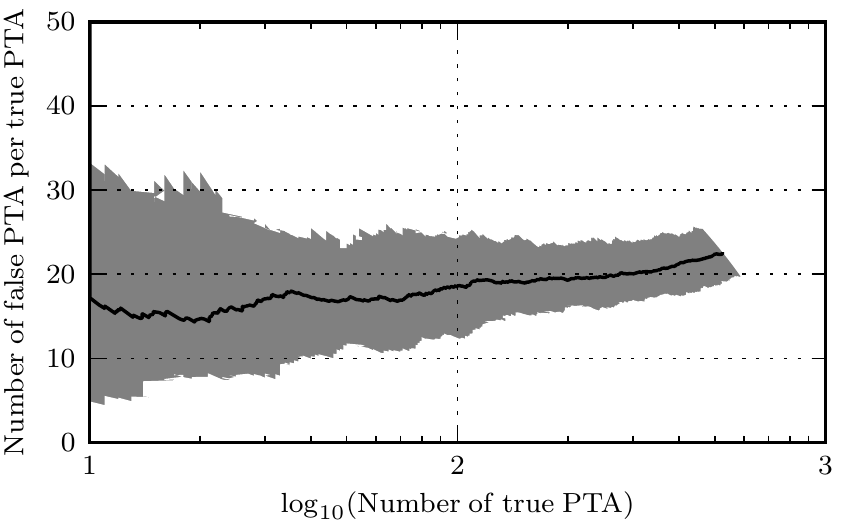}
\caption{Efficiency of the search for PTA-galaxies in the adapted catalog.
This plot is analogous to that in Figure \ref{fig:threshold_bgal}, but now using the probabilities $p_a(\text{B,P}|\theta_i,\mathcal{M})$.
Hence, galaxies in this plot are selected such that their hosted MBH produces (if it is a binary) a strain amplitude larger than the threshold $h_0^\text{thres}=10^{-15}$.}
\label{fig:threshold_pta}
\end{figure}

\begin{figure}
\includegraphics{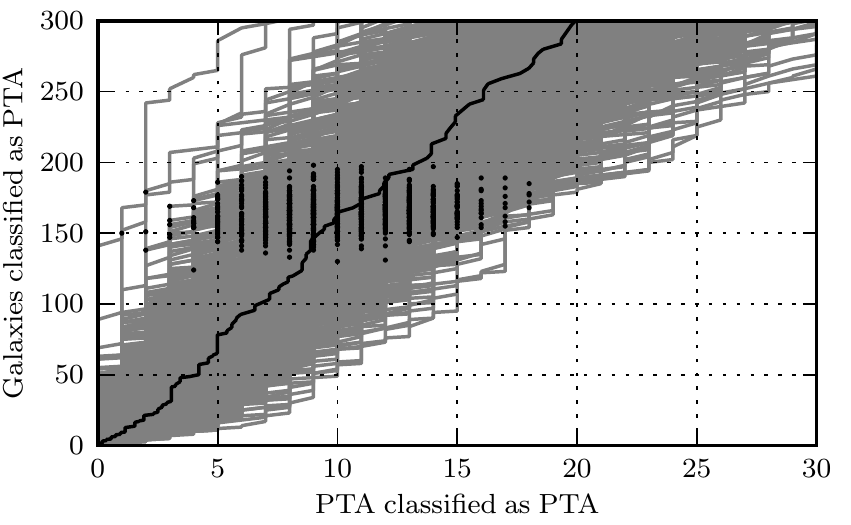}
\caption{Total number of galaxies classified as PTA-galaxies versus the number of good classifications.
Each gray curve corresponds to a different realization (a different choice of galaxies from the adapted catalog covering $19.5\%$ of the sky).
The black line is the average over all 1000 realizations.
For each realization, we plot a black point corresponding to the PTA-galaxy probability threshold $p_T^\text{PTA}$.
Galaxies considered here are such that their hosted MBH produces (if it is a binary) a strain amplitude larger than the threshold $h_0^\text{thres}=10^{-15}$.}
\label{fig:threshold_all}
\end{figure}

Figure \ref{fig:threshold_pta} is analogous to \ref{fig:threshold_bgal}, but now the PTA-galaxy probabilities are used, instead of the B-galaxy probabilities, to test the efficiency of the search in the adapted catalog.
This plot is used to decide upon a probability threshold $p_T^\text{PTA}$ to select our real PTA-galaxy candidates later on.
Let us first consider the case in which the chosen probability threshold corresponds to the rightmost point on the plot.
The worst results for such a threshold find 462 PTA-galaxies among 11764 candidates, whereas the best results have 588 PTA-galaxies of 11610 candidates.
The average is a number of 525 PTA-galaxies of a total of 11770 candidates.
An important remark is that these $\sim500$ PTA-galaxies are all B-galaxies with $z<0.1$ that produce a strain amplitude larger than $10^{-15}$ in the adapted catalog.
In other words, \textit{all} possible PTA-galaxies contained in the adapted catalog are counted as candidates, by choosing such a probability threshold.

A detection rate of 100$\%$ is a great feature, however, the false alarm rate is large.
The list of candidates can be reduced, by choosing a larger probability threshold, so that the number of bad candidates per good one decreases (by the cost of losing PTA-galaxies).
We now want to select a threshold such that only the $\sim 10$ most likely PTA-galaxies are counted as candidates.
For that, we choose a probability threshold corresponding to the leftmost point in Figure \ref{fig:threshold_pta}.
This threshold is $p_T^\text{PTA}=8.01\times 10^{-06}$.
Now, in the worst realizations, 1 PTA-galaxy is found among 190 candidates; whereas in the best realizations 18 PTA-galaxies are counted among 137 candidates.
On average, we count 10 PTA-galaxies in a list of 164 candidates.

In Figure \ref{fig:threshold_all} we have plotted the same data as in Figure \ref{fig:threshold_pta} in a different manner (and at a different range of probabilities).
Each gray line corresponds to one realization; the black line contains the average values over all (1000) realizations.
Each one of the 1000 black points gives the number of galaxies (on the vertical axis) and PTA-galaxies (on the horizontal axis), for a particular realization, that have PTA-galaxy probabilities larger than $p_T^\text{PTA}$.
Here we see that, for such a threshold, $\sim 10$ galaxies among the $\sim150$ selected as candidates are well classified.
We expect similar numbers when applying the search to the real catalog.

\subsection{Efficiency of the search when including clustering information}
\label{subsec:search_clustering}
\begin{figure}
\includegraphics{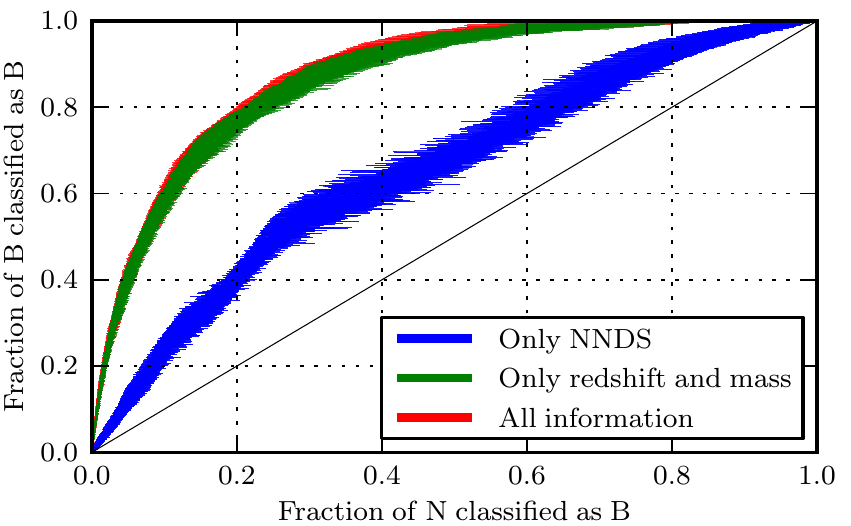}
\includegraphics{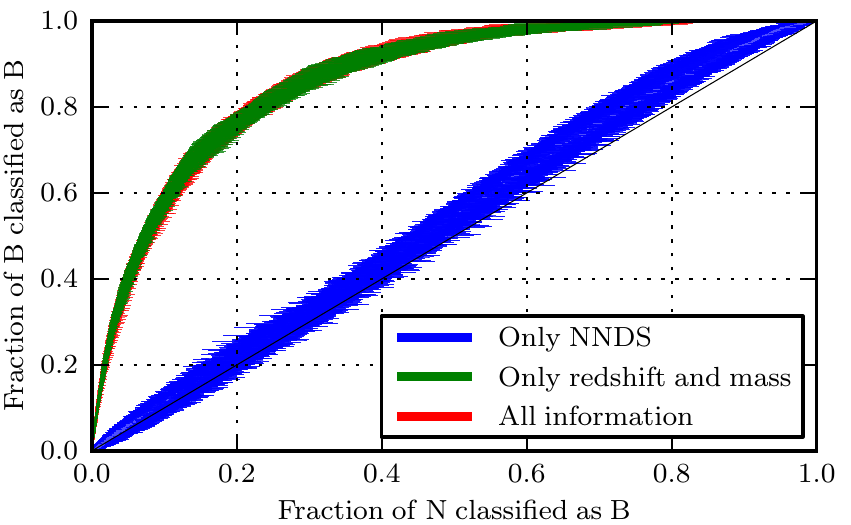}
\caption{ROC curves obtained for the search for B-galaxies using a MLA.
The upper and lower plots correspond to case (i) and (ii) (described in Section \ref{subsec:nnds}), respectively.
Filled areas contain all the curves produced after the 1000 realizations.
Blue and green curves correspond to searches in which only the NNDS, and only $z$-$m$, respectively, are taken as features.
Red curves are the results of searches in which all parameters ($z$, $m$, and NNDS) are considered.}
\label{fig:ROC_clustering}
\end{figure}

We now turn to the results of the search including clustering information, as described in Section \ref{subsec:nnds}.
We applied the MLA 1000 times; for each realization, the systems of the training set were different, but the testing set was always composed of the galaxies of one particular patch of the sky (of the 9 into which the adapted catalog is divided).
After each realization we construct a ROC curve with the probabilities assigned to the B- and N-galaxies of the testing set.
In order to determine the amount of information provided by the parameters ($z$, $m$ and NNDS), the MLA is used in three different circumstances.
First, we characterize galaxies only using the NNDS (blue curves in the plots of Figure \ref{fig:ROC_clustering}).
Second, only $z$ and $m$ are used as features (green curves).
Third, all pieces of information ($z$, $m$, and NNDS) are considered (red curves).
The NNDS is always calculated for 50 shells with borders separated from 0 to 4\,Mpc in linear steps.

Figure \ref{fig:ROC_clustering} shows the ROC curves resulting from this search.
Upper and lower plots correspond to cases (i) and (ii) (as described in Section \ref{subsec:nnds}), respectively.
The colored areas contain all ROC curves after the 1000 realizations.
In the lower plot, the NNDS is affected by the inclusion of peculiar velocities and by the subtraction of neighbors; the efficiency of the search using only the NNDS is not significantly different from a totally random search.
A discouraging result is that the inclusion of the NNDS introduces some level of randomness in the search.
In the upper plot of Figure \ref{fig:ROC_clustering}, we see that the red and green areas are approximately equal; but in the lower plot, the red area is shifted to the right (to higher values of false alarm rate) with respect to the green one.

Summarizing, the inclusion of the clustering does not ameliorate (in case (i)) or even deteriorates (in case (ii)) the efficiency of the search.
However, the fact that the blue curve differs from the black thin line, in the upper plot, means that the clustering of B-galaxies presents in fact features that distinguish them from N-galaxies.
The plots in Figure \ref{fig:ROC_clustering} change, although not drastically, by using other choices of the shells, reducing the number of shells (from 50 to 10 or 20), or assigning the testing set to another patch of the sky.
Nevertheless, the conclusions already drawn about the efficiency of the searches using the NNDS remain unaffected.
By comparing the ROC curves obtained in this and in Section \ref{subsec:efficiency}, one can also conclude that the efficiency of our Bayesian search is consistent to that of the MLA (when characterizing galaxies only by their redshift and mass).

\subsection{Efficiency of the search at larger redshifts}
\label{subsec:search_extending}
\begin{figure}
\includegraphics{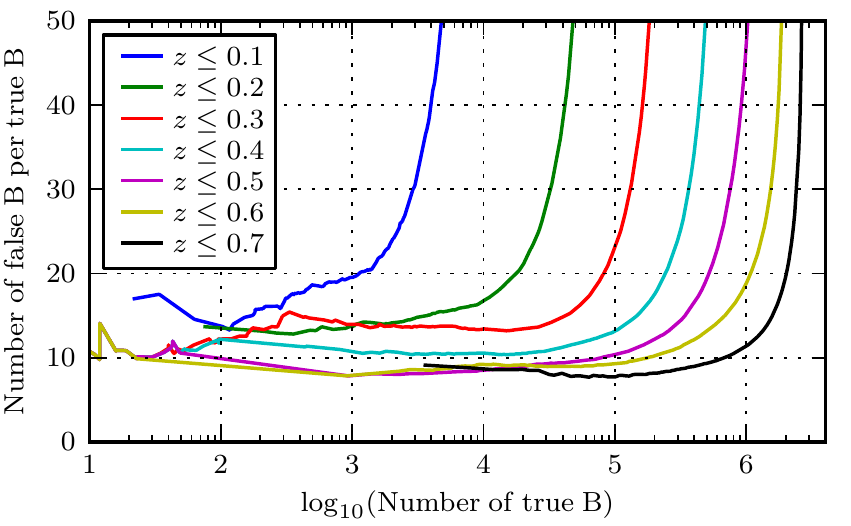}
\includegraphics{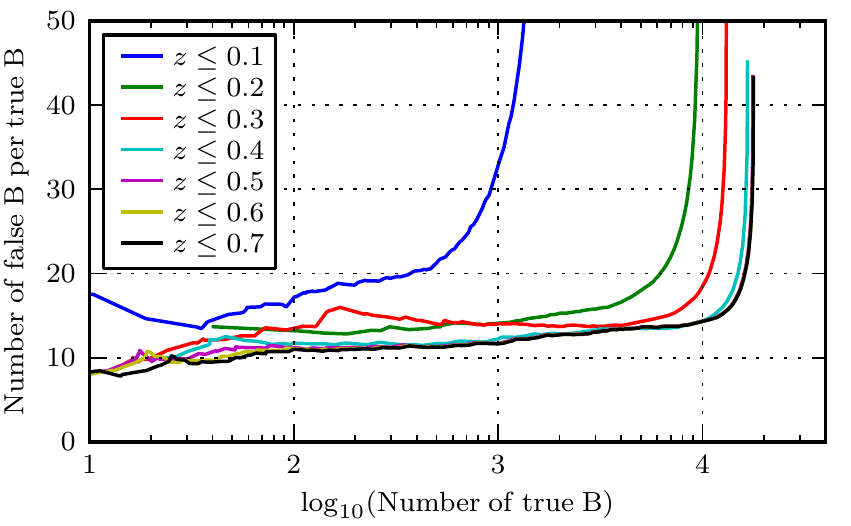}
\caption{Estimate of the efficiency of a search for B-galaxies extended to larger redshifts (up to 0.7).
The plot shows the ratio of bad classifications per good one versus good classifications of B-galaxies, within different redshift intervals.
The upper (or lower) plot is obtained when applying the extended search to the fake (or adapted) catalog.}
\label{fig:threshold_larger_z}
\end{figure}

Here we show how well the search for B-galaxies performs at redshifts as large as 0.7.
The probabilities $p_x(\text{B}|\theta_i)$ are calculated following the method described in Section \ref{subsec:extending}.
We build a $z$-$m$-histogram of $100\times100$ pixels; redshifts and masses are in the ranges $[z_\text{min},z_\text{max}]$ and $[m_\text{min},m_\text{max}]$, respectively.
Each pixel is assigned a value of $p_x(\text{B}|\theta_i)$.
We then count the number of B- and N-galaxies contained in pixels with probabilities larger than a certain value that goes from 0 to 1.
We do this for different maximum redshifts: 0.1, 0.2, $\dots$, 0.7.
The results are depicted in Figure \ref{fig:threshold_larger_z}, that contains two plots similar to those in Figures \ref{fig:threshold_bgal} and \ref{fig:threshold_pta}.
Upper and lower plots correspond to the probabilities $p_f(\text{B}|\theta_i)$ (using the fake catalog) and $p_a(\text{B}|\theta_i)$ (using the adapted catalog), respectively.

One notices with Figure \ref{fig:threshold_larger_z} that the number of bad classifications per good one does not grow (and even decreases) when considering galaxies at larger redshifts. 
For example, considering systems in the adapted catalog all the way up to $z_\text{max}$, there exists a certain probability threshold such that $\sim 1000$ B-galaxies (and a factor of $\sim 11$ more N-galaxies) have larger probabilities than that threshold.
Applying this search to the real catalog we could then make a list with $\sim 11000$ candidates containing $\sim 1000$ true B-galaxies.
If the trend found on the search for PTA-galaxies at $z<0.1$ (in Section \ref{subsec:efficiency_pta}) also holds at larger redshifts, we could then presumably make a list with the, say, $\sim 1000$ most likely single PTA sources contained in the SDSS footprint (for $z<0.7$).

Although extending the searches to larger redshifts looks potentially very interesting, such a task is not performed for the moment.
We believe that, to create a trustworthy list of real PTA-galaxy candidates at such large redshifts, the method used to adapt our fake catalog to the limitations of the real one should be more accurate than that described in Section \ref{subsec:extending}.
A proper extension of the search towards larger redshift is thus left for a possible follow-up work.

\subsection{Assigning probabilities to real galaxies}
\label{subsec:real_galaxies}

\begin{figure*}
\includegraphics{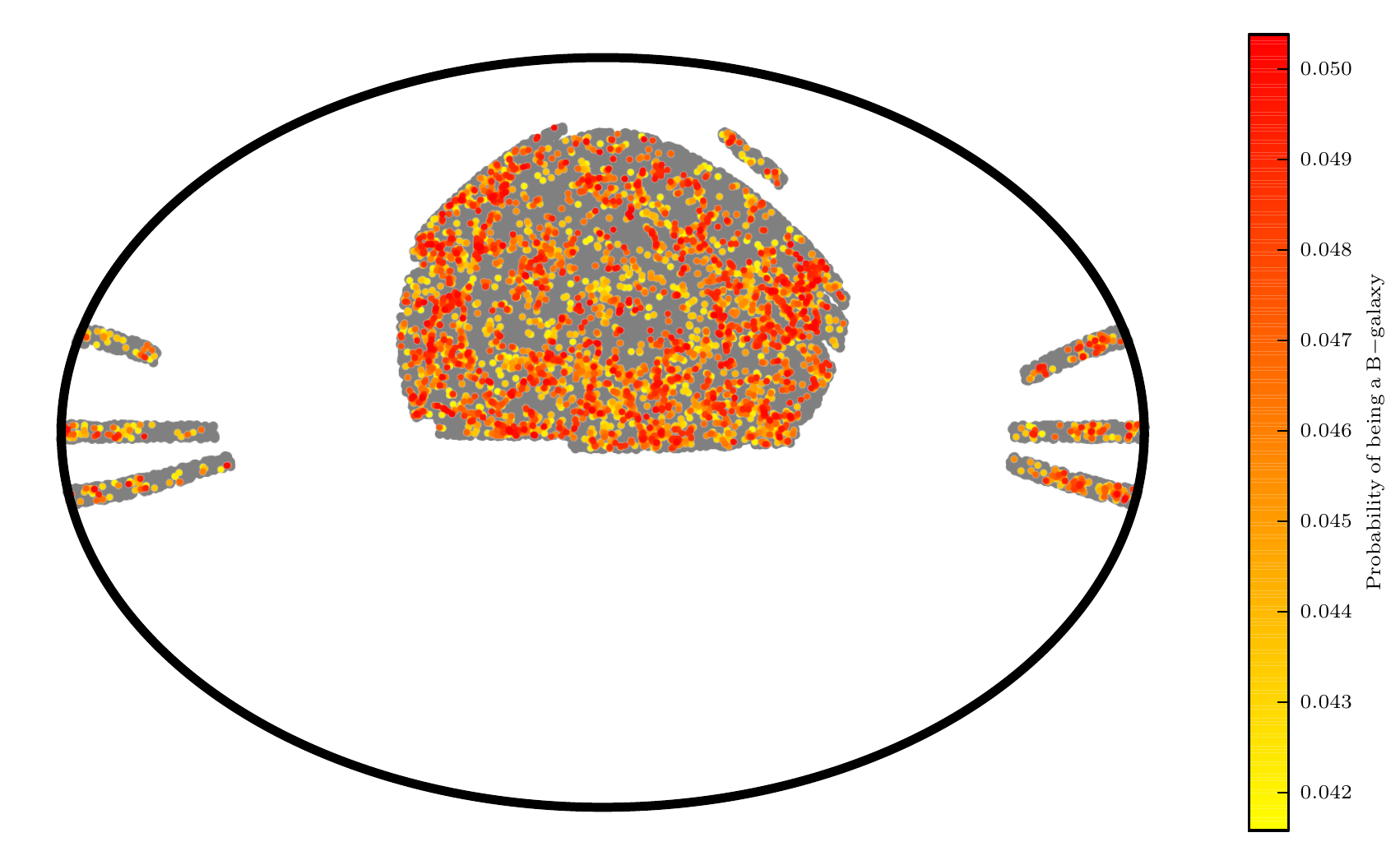}
\caption{Projected skymap (using a Hammer projection) of galaxies from the real catalog.
Colored points are B-galaxies candidates.
The color bar gives the B-galaxy probability $p_r(\text{B}|\theta_i)$; redder points are galaxies with larger probabilities.
Gray points are galaxies with probabilities below the chosen threshold $p_T^\text{B}$.
From the 3870 candidates we expect $\sim 196$ of them to be real B-galaxies.}
\label{fig:skymap_bgal}
\end{figure*}

A B-galaxy probability is now assigned to each system in the real catalog.
This probability, determined by the matrix $p_r(\text{B}|\theta_i)$ (whose construction is explained in Section \ref{subsec:bayesian_search}), depends only on the bin $\theta_i$ in which the values of $z$ and $m$ of the galaxy are contained.
In Figure \ref{fig:skymap_bgal} we plot a projected skymap of the systems from the real catalog that have probabilities larger than the threshold $p_T^\text{B}$ introduced in Section \ref{subsec:efficiency}.
All galaxies that are not candidates are gray points; B-galaxy candidates have colors corresponding to different values of probabilities $p_r(\text{B}|\theta_i)$.
Redder points are candidates with larger B-galaxy probabilities.
In this skymap there are 3870 candidates, a factor $\sim 1.79$ more systems than in the adapted catalog for the same probability threshold $p_T^\text{B}$.
The reason for this is the overabundance of high-mass galaxies in the SDSS with respect to the MS, discussed in Section \ref{sec:discussion}.
We thus expect that, among these candidates, $\sim 1.79\times 110\approx 196$ are true B-galaxies (since 110 was the number of B-galaxies found in the adapted catalog for the same probability threshold).

In order to assign to each real galaxy its probability of hosting a MBHB observable by the PTA, $p_r(\text{B},\text{P}|\theta_i,\mathcal{M})$, we follow the prescriptions described in Section \ref{subsec:search_pta}; for that we first need the bulge mass $m_\text{bulge}$ of each galaxy,
\begin{equation}
\label{eq:mbulge}
m_\text{bulge}=f_b m,
\end{equation}
where $f_b$ is the bulge mass fraction.
Elliptical galaxies are expected to have $f_b$ close to 1, while for spiral galaxies, reasonable values of $f_b$ lie between $\sim 0.1$ and $\sim0.3$.
We now explain how these bulge mass fractions have been obtained.
The Galaxy Zoo \citep{LintottEtAl2008,LintottEtAl2011} is a project in which volunteers assign SDSS galaxies a morphological classification by visual criterion.
The data are public\footnote{\url{http://data.galaxyzoo.org}} and contain a final morphological-type classification, constructed after processing the votes of the volunteers and reducing possible visual biases.
The three possible types are ``elliptical'', ``spiral'', and ``unknown'' (in those cases when the voting for elliptical or spiral was not significant enough).
Unfortunately, not all galaxies in our real catalog have a final classification in the Galaxy Zoo.
We use those galaxies with a type classification different than ``unknown'' that are contained in both catalogs, to adopt a criterion on the morphologies of all galaxies in our real catalog.

\begin{figure}
\includegraphics{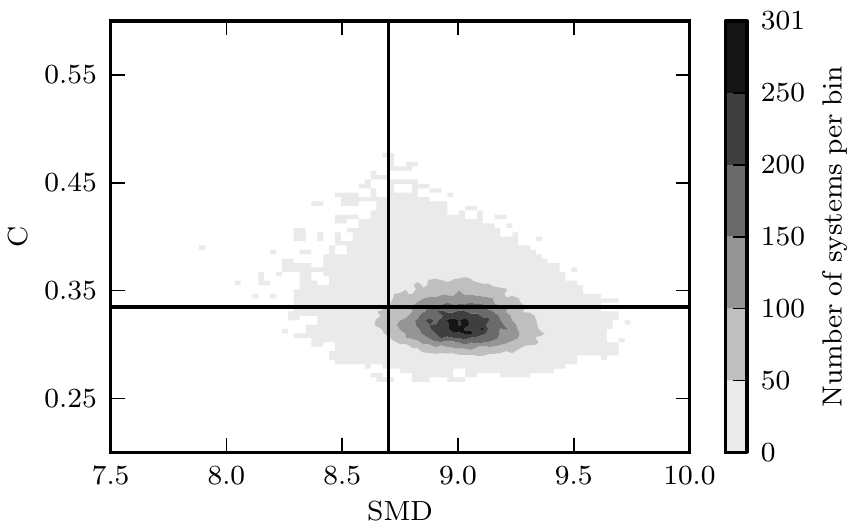}
\includegraphics{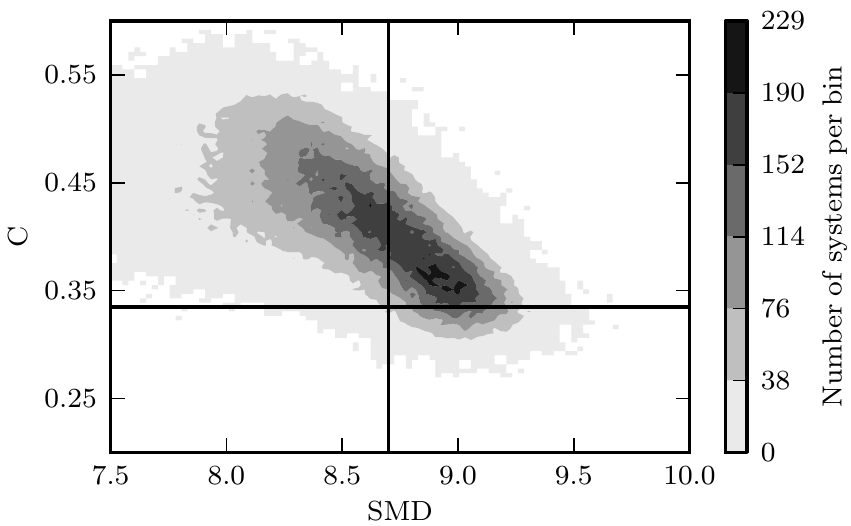}
\caption{Contour plots of the distribution of galaxies classified as elliptical (above) and spiral (below) in the Galaxy Zoo.
The horizontal axis shows the surface mass density, and the vertical one is the concentration parameter.
Both axes are divided into 100 linearly spaced bins, and the color bar gives the number of systems per bin.
We use these distributions to find a criterion to calculate the bulge mass fractions of systems in our real catalog.}
\label{fig:morpho}
\end{figure}

\begin{figure}
\includegraphics{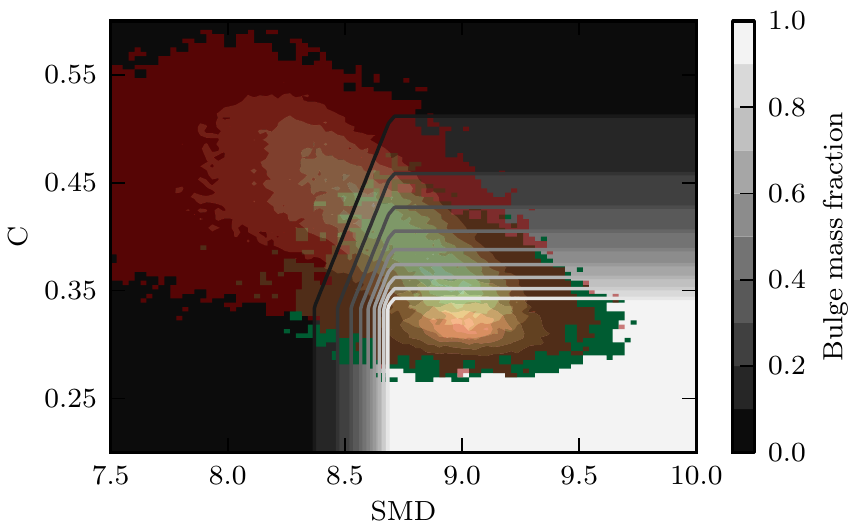}
\caption{Contour plot of the function $f_b(\text{SMD},C)$, that gives, at each point of the SMD-$C$ plane, the bulge mass fraction that we assign to galaxies in our real catalog.
Galaxies falling within the lower-right square (with a white background) will have a bulge mass fraction equal to 1, whereas the bulge mass fraction of galaxies far from these region will decay exponentially.
On top of it, the contour plots of Figure \ref{fig:morpho} have been superimposed: elliptical galaxies from the Galaxy Zoo on green and spiral ones on red.}
\label{fig:mbh_morpho}
\end{figure}

In Figure \ref{fig:morpho} we show contour plots of galaxies classified in the Galaxy Zoo as ellipticals (on top) and as spirals (bottom).
The horizontal axis shows the \textit{surface mass density} (SMD), defined as
\begin{equation}
\text{SMD}=\log_{10}\left( \frac{m}{2\pi R^2} \frac{\text{kpc}^2}{\msun}\right).
\end{equation}
The radius $R$ is the half-light proper radius in the z-band, calculated using
\begin{equation}
R=[1+z]^{-1}r(z)\mathcal{R}_{50,\text{z}},
\end{equation}
where $r(z)$ is the comoving distance to the galaxy (Equation (\ref{eq:comoving})), and $\mathcal{R}_{50,\text{z}}$ is the angular radius in which $50\%$ of the Petrosian flux in the z-band is contained (called \verb|petroR50_z| in the SDSS server).
The vertical axis shows the concentration parameter $C$, for which we use the definition
\begin{equation}
C=\frac{\mathcal{R}_{50,\text{r}}}{\mathcal{R}_{90,\text{r}}}.
\end{equation}
This means, $C$ is the ratio of the $50\%$ and $90\%$ Petrosian radii in the r-band (called \verb|petroR50_r| and \verb|petroR90_r| in the SDSS server, respectively).
The query used to obtain the parameters \verb|petroR50_z|, \verb|petroR50_r|, and \verb|petroR90_r| is included in Appendix \ref{sec:queries}.
From Figure \ref{fig:morpho} we see that elliptical and spiral galaxies cannot be clearly distinguished in a certain region of the SMD-$C$ plane; there is an important overlap.
The darkest regions in the upper and lower plots show the maximum accumulation of elliptical and spiral galaxies, respectively.
The intermediate value of $C$ between those two maxima is $\overline{C}=0.335$ (which is plotted as a black horizontal line).
The vertical line is at the value of SMD above which $90\%$ of elliptical galaxies are contained, $\overline{\text{SMD}}=8.70$.
We take these two values and construct the functions
\begin{equation}
f_b^\text{SMD}(c_s,\overline{\text{SMD}},\text{SMD})=\min \left( 1, \exp \left( c_s [\text{SMD}-\overline{\text{SMD}}]\right) \right)
\end{equation}
and
\begin{equation}
f_b^C(c_r,\overline{C},C)=\min \left( 1, \exp \left( c_r [C-\overline{C}]\right) \right).
\end{equation}
Then, the bulge mass fraction is calculated as the product of the two previous functions,
\begin{equation}
\label{eq:fbfun}
f_b(\text{SMD},C)=f_b^\text{SMD}(7,8.70,\text{SMD})f_b^C(-13,0.335,C),
\end{equation}
where the parameters $c_s$ and $c_r$ have been chosen in such a way that the average $f_b$ is $\sim 0.9$ for elliptical galaxies and $\sim 0.3$ for spiral galaxies.
We point out that this particular choice of functions and parameters is arbitrary; the aim of this calculation is to construct a simple procedure to assign bulge mass fractions based on observational data.
In Figure \ref{fig:mbh_morpho}, a contour plot of $f_b(\text{SMD},C)$ is shown, and on top we have superimposed the distributions of ellipticals (on green) and spirals (on red) previously shown (in Figure \ref{fig:morpho}).

\begin{figure*}
\includegraphics{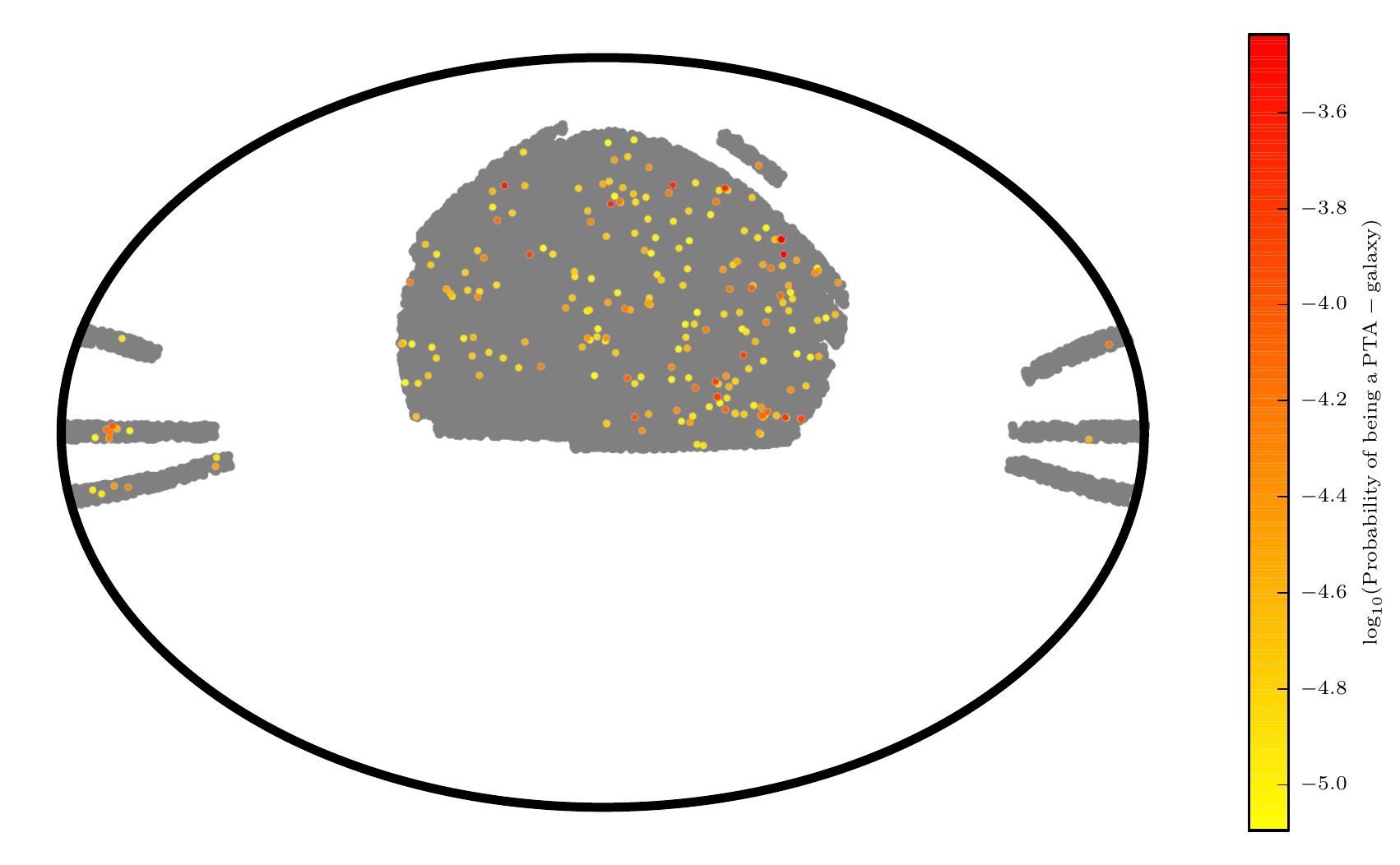}
\caption{Projected skymap of galaxies from the real catalog (analogous to Figure \ref{fig:skymap_bgal}).
Now the color bar gives the PTA-galaxy probabilities.
Colored points in this plot correspond to real PTA-galaxy candidates, i.e. galaxies that may host a MBHB emitting GWs that produce a maximum strain amplitude larger than $h_0^\text{thres}=10^{-15}$.
We expect $\sim 14$ of these 232 candidates to actually be B-galaxies (so they may contain an observable MBHB).
The probability of observing one of them is small (as the numbers in the color bar reveal); however, if we do observe a single source from this region of the sky (at $z<0.1$) with a sensitivity of $h_0^\text{thres}$, it will most likely be one of these candidates.}
\label{fig:skymap_pta}
\end{figure*}

Once the bulge masses are known, the rest of the calculation of the PTA-galaxy probabilities is as described in Section \ref{subsec:search_pta}.
The list of real PTA-galaxy candidates is constructed as explained in Section \ref{subsec:efficiency_pta}: we select galaxies from the real catalog that have PTA-galaxy probabilities larger than $p_T^\text{PTA}$.
In Figure \ref{fig:skymap_pta} we show a projected skymap with these candidates.
They are 232, which is a factor of $\sim 1.41$ larger than the average number of candidates produced in the adapted catalog.
Therefore, we expect that $\sim 1.41 \times 10\approx 14$ true PTA-galaxies are counted among them.
These candidates are the most likely single PTA sources in the local universe (contained in the SDSS window at $z<0.1$).
Note that we do not expect to observe $\sim 14$ MBHBs emitting at $h_0 \ge 10^{-15} $ in the SDSS window; in fact, the PTA-galaxy probability of each of the candidates is fairly small (see the numbers in the color bar of Figure \ref{fig:skymap_pta}), since they spend a relatively short interval of time emitting at frequencies at which they are observable.
However, if we do observe a PTA source from this part of the sky with $z<0.1$, it will most likely be one of these candidates.
If one wanted to perform a targeted search, one could use this list of galaxies; the list could also be reduced by combining ours with other searching criteria, for example, by looking for signs of recent galaxy interaction in the SDSS images (that can be accessed from the SDSS server).

As we pointed out in Section \ref{subsec:efficiency_pta}, we could also construct a list of PTA-galaxy candidates such that \textit{all} possible PTA-galaxies observed by the SDSS are counted.
Taking into account this factor of $\sim 1.41$ difference with respect to the adapted catalog, the resulting list would contain $\sim 1.41 \times 11770 \approx 1.66\times 10^{4}$ galaxies, of which $\sim 1.41 \times 525 \approx 740$ would be PTA-galaxies.
They would be the only galaxies in the local universe that could possibly be observed emitting GWs of a strain amplitude $h_0 \ge 10^{-15}$ within the spectroscopic SDSS catalog.

\begin{figure}
\includegraphics{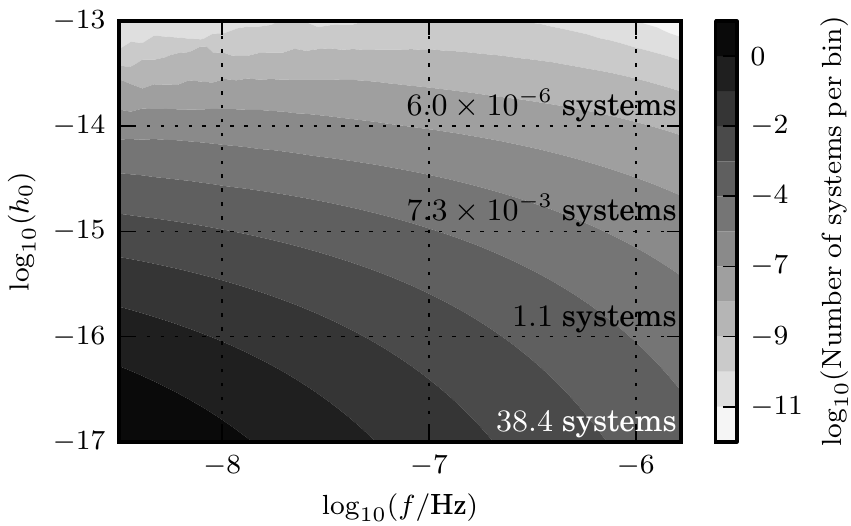}
\includegraphics{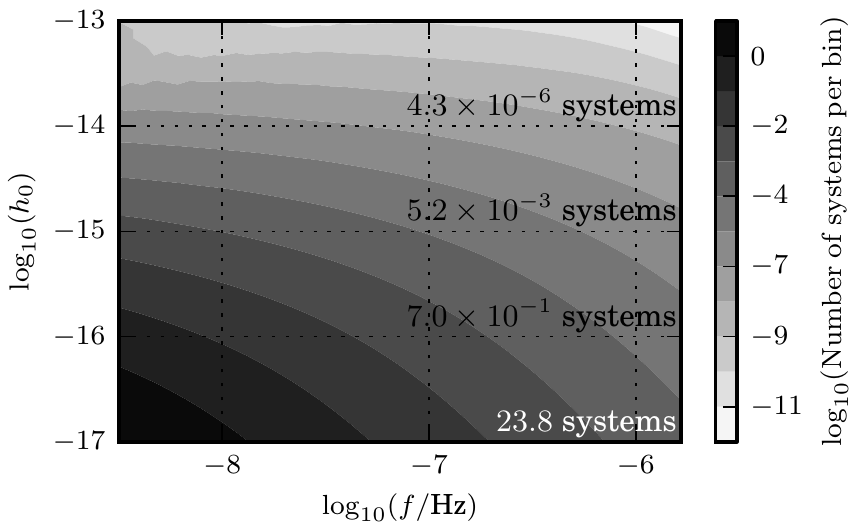}
\caption{Strain amplitude versus observed GW frequency (averaged over 100 realizations).
The horizontal axis is divided into 50 frequency bins, and the vertical axis into 50 strain amplitude bins.
The color of each pixel in the upper plot gives the sum of probabilities of all galaxies from the real catalog (with $z<0.1$) that produce a strain amplitude larger than a certain value (given in the vertical axis) within a certain frequency bin (given in the horizontal axis). 
The numbers written on top give the sum of probabilities over all frequency bins at a fixed strain amplitude, i.e. these are the average numbers of PTA-galaxies (for different strain amplitude thresholds) that are contained in the real catalog at $z<0.1$.
The lower plot is analogous to the upper one, but for systems in the adapted catalog.}
\label{fig:h0_vs_f}
\end{figure}

\begin{figure}
\includegraphics{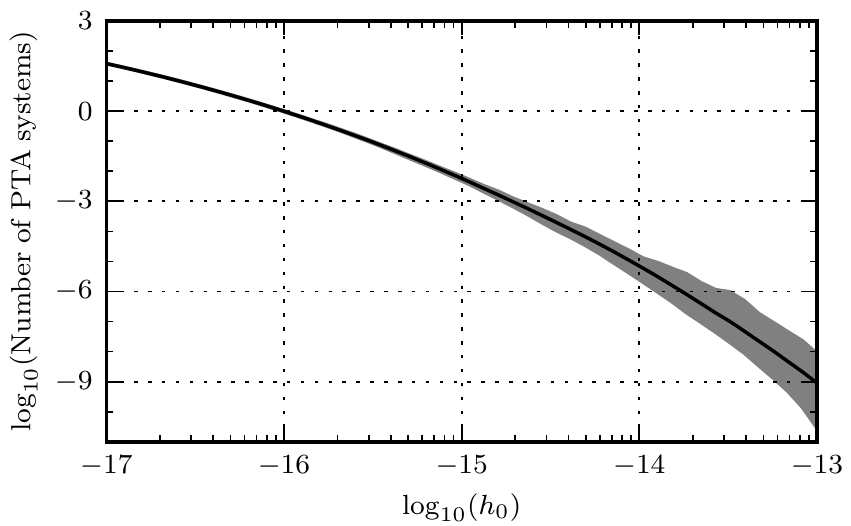}
\includegraphics{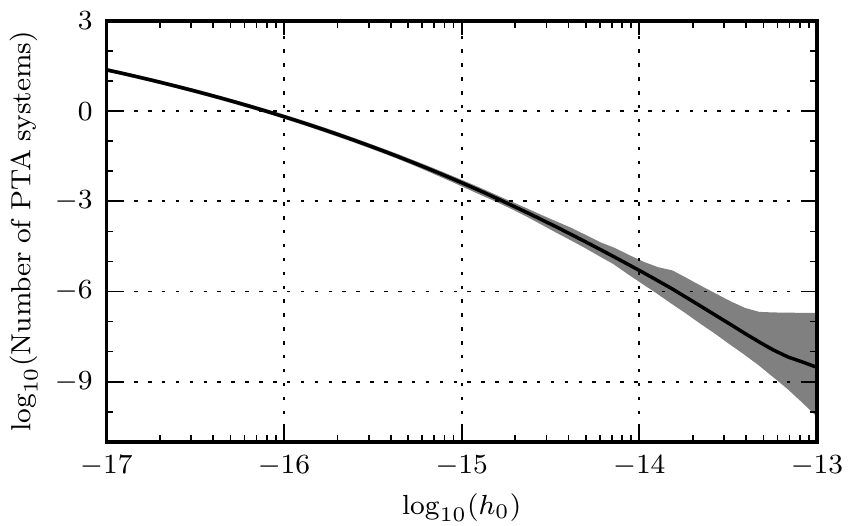}
\caption{Average number of PTA-galaxies that are contained in the real (above) and adapted (below) catalog for $z<0.1$, as a function of the strain amplitude threshold.
For example, to have an average of 1 galaxy in the real catalog, the PTA needs to be able to detect strain amplitudes smaller than $10^{-16}$.
These plots are obtained by integrating the plots in Figure \ref{fig:h0_vs_f} over all frequencies in the PTA frequency window.}
\label{fig:numsys}
\end{figure}

In Figure \ref{fig:h0_vs_f} the strain amplitude of systems in the real catalog is plotted as a function of the observed GW frequency.
Each point of the upper plot gives the average number of galaxies from the real catalog with $z<0.1$ that can be found within a certain frequency bin (whose central frequency is given by the horizontal axis), producing a strain amplitude larger than a certain value (given by the vertical axis).
The numbers written on top of the graph are the sum over all frequency bins of the window, for a particular strain amplitude threshold ($10^{-17}$, $10^{-16}$, $10^{-15}$, and $10^{-14}$).
For example, we expect on average 1.1 systems in the real catalog producing a strain amplitude larger than $10^{-16}$ within the PTA frequency window.
The number of systems is calculated as the sum of the probabilities $p_r(\text{B},\text{P}|\theta_i,\mathcal{M})$ of all galaxies.
The lower plot in Figure \ref{fig:h0_vs_f} represents the same as the upper one, but for systems in the adapted catalog.
We see that the numbers in both plots agree well; although the numbers are slightly smaller in the adapted catalog case, again due to the shortage of systems at the high-mass end, with respect to the real catalog (see Section \ref{sec:discussion}).
In Figure \ref{fig:numsys} the total number of PTA-galaxies expected to be observed with $z<0.1$ is plotted against the strain amplitude threshold.
The gray area contains the curves obtained for each of the 100 realizations described in Section \ref{subsec:search_pta}.
The black line is the average over all realizations.
Upper and lower plots count systems from the real and adapted catalogs, respectively.

\begin{figure}
\includegraphics{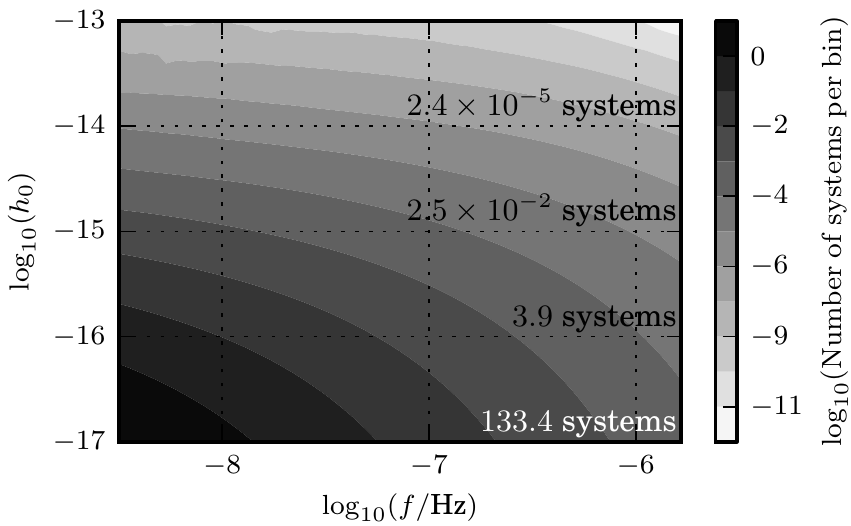}
\includegraphics{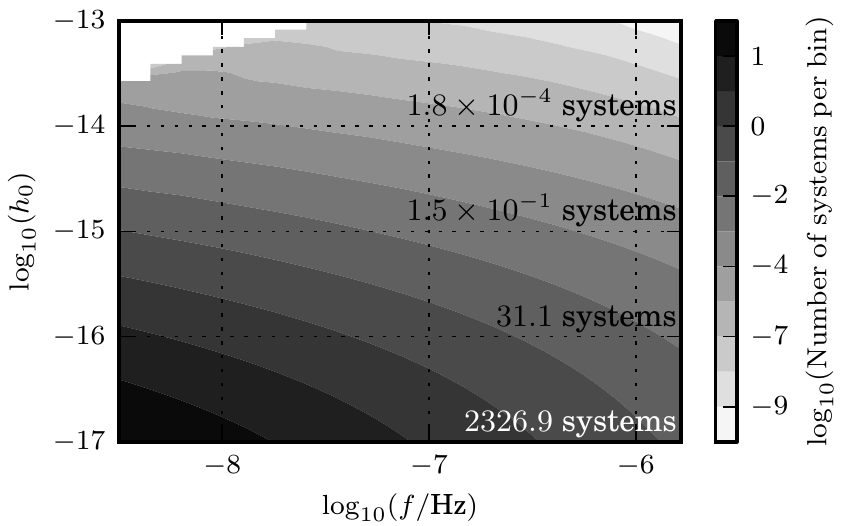}
\caption{The upper plot is analogous to the upper plot in Figure \ref{fig:h0_vs_f}; now, all systems in the real catalog (with $z<0.7$) are considered.
The lower plot is calculated considering all B-galaxies from the entire fake catalog; on top are written the average numbers of PTA-galaxies (for different strain amplitude thresholds) that could be observed in a very complete, all-sky galaxy catalog (with $z<0.7$).
The upper-left corner is empty simply because, among the 100 realizations, there were no B-galaxies observable at that region of frequencies producing such a large strain.}
\label{fig:h0_vs_f_larger_z}
\end{figure}

\begin{figure}
\includegraphics{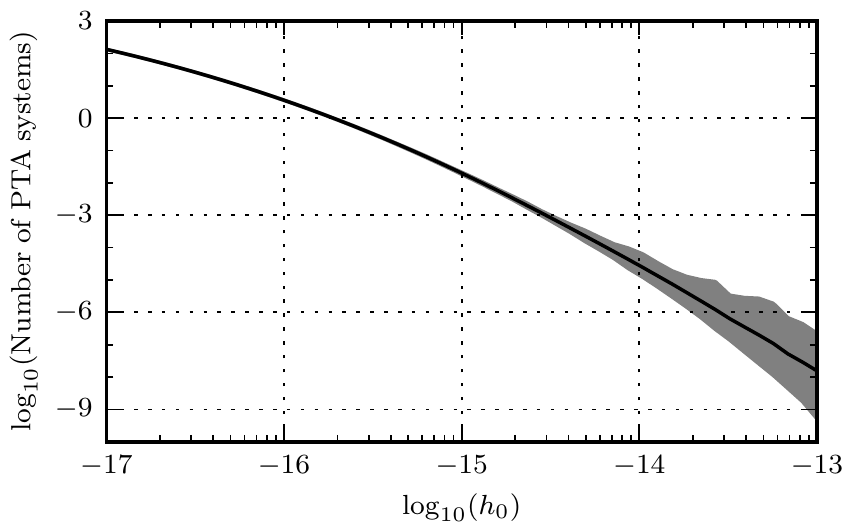}
\includegraphics{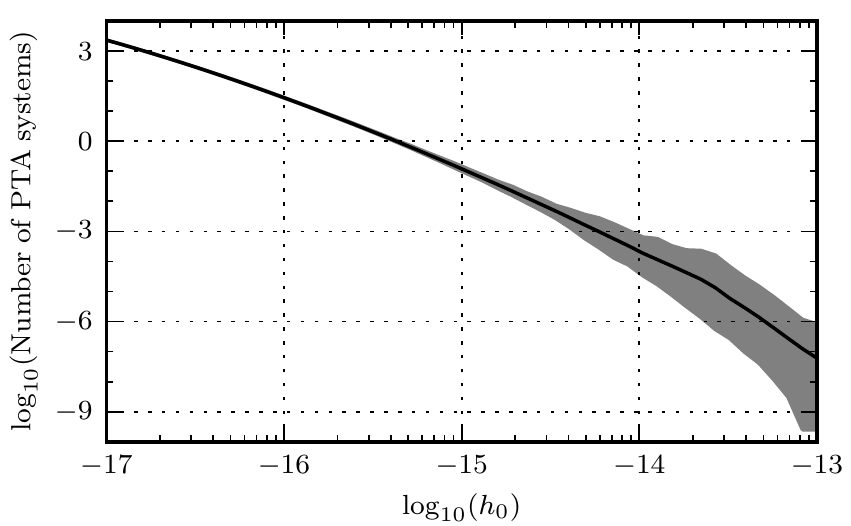}
\caption{The upper plot is analogous to the upper plot in Figure \ref{fig:numsys}, but now all systems in the real catalog are considered (with $z<0.7$).
The lower plot gives the average number of PTA-galaxies that would be contained in an very complete, all-sky galaxy catalog (the fake catalog) as a function of their maximum strain amplitude.}
\label{fig:numsys_larger_z}
\end{figure}

The upper plot in Figure \ref{fig:h0_vs_f_larger_z} is analogous to the upper plot in \ref{fig:h0_vs_f}, but now considering systems with $z<0.7$.
Still, in order to count an average number of PTA-galaxies larger than 1 in the real catalog, a strain amplitude threshold smaller than $\sim 10^{-16}$ is necessary.
The lower plot in Figure \ref{fig:h0_vs_f_larger_z} shows the same as the upper one, but considering only B-galaxies from the whole fake catalog.
In this case, the probabilities $p_f(\text{B},\text{P}|\theta_i,\mathcal{M},f)$ are also defined by Equation (\ref{eq:pfbp}), but the B-galaxy probabilities $p_f(\text{B}|\theta_i)$ are identically 1 for all galaxies (since we know they are B-galaxies).
Hence, this graph reveals the average number of PTA-galaxies that would be contained in an ideal all-sky galaxy catalog (more complete than a spectroscopic catalog like the one we use).
One could plot, on top of the graphs in Figure \ref{fig:h0_vs_f_larger_z}, the exact sensitivity of the PTA to single MBHBs for a given array of MSPs \citep{EllisEtAl2012}; the sum of the pixels swept by the sensitivity curve would give the average number of systems that should be observable for such an array.
The total number of PTA-galaxies with $z<0.7$ as a function of the strain amplitude threshold is shown in Figure \ref{fig:numsys_larger_z}, for galaxies of the real catalog (upper plot) and for B-galaxies of the entire fake catalog (lower plot).
The upper plot is analogous to the upper plot in Figure \ref{fig:numsys}, but now considering all systems in the real catalog, and not only those with $z<0.1$.
The lower graph informs on the average number of PTA-galaxies that could be observed with an ideally complete all-sky galaxy catalog.

\section{Caveats and assumptions}
\label{sec:discussion}

Our selection of B-galaxies is based on the assumption that MBHBs form and coalesce in less than $\sim$300\,Myr, which is the approximate time between MS snapshots (at least for $z<0.7$).
In practice, in the MS a galaxy merger starts when the dark matter subhalo of the satellite galaxy is disrupted, and ends after a dynamical friction time computed according to Equation (31) of \cite{GuoEtAl2011b} \citep[from][]{BinneyTremaine2008}.
It is at this moment that the system is recorded as `merger'; however, the two MBHs are likely still orbiting each other at a separation of few parsecs, because dynamical friction becomes inefficient on them once they have paired in a Keplerian binary.

Let us suppose that a merger occurred between two consecutive snapshots, $n-1$ and $n$.
This merger is first registered by the simulation at snapshot $n$.
At this moment, the descendant galaxy has three possibilities: (i) it contains a single MBH (because the MBHB has already coalesced between $n-1$ and $n$), (ii) it contains a wide MBHB (that is not yet emitting strong enough GWs to be observed), (iii) it contains a MBHB that is emitting observable GWs (and will coalesce well before $n+1$).
Ideally many B-galaxies would be in (iii), however, this is in reality the least likely case; the time a MBHB emits GWs that produce a strain on Earth larger than, for instance, $10^{-15}$, is usually much smaller than the time between snapshots\footnote{This is the reason why the PTA-probabilities calculated in Section \ref{subsec:search_pta} are so small.}.
Hence, we do not expect to have many B-galaxies in (iii), but we do assume the most of the coalescences to occur closer to $n$ than either $n-1$ or $n+1$, i.e., that the binary lifetime is less than $\sim300\,$Myr.
This assumption is supported by recent simulations of MBHB mergers in stellar bulges \citep{PretoEtAl2011,KhanEtAl2012}, which found that the `hardening phase' (i.e. the lifetime) of most MBHBs would last at most a few hundred Myr.
While the role and impact of cold gas on such massive systems is less clear, any additional source of friction would help making this time-scale shorter.

Suppose now that the mechanisms taking part after the dynamical friction evolve over time-scales larger than $\sim300\,$Myr.
Then, the systems selected in $n$ as B-galaxies should be those galaxies that merged between $n-2$ and $n-1$, or even at earlier snapshots, in order to account for the delay.
This alternative selection would introduce a displacement in the redshift distribution of B-galaxies.
However, as Figures \ref{fig:ROC_MS_bayesian} and \ref{fig:ROC_adapted_bayesian} show, redshifts do not ameliorate much the search for B-galaxies, so the displacement would not change our results considerably.
The discriminating parameter in the search is the mass, which is not expected to vary significantly between two snapshots after the merger.
As a conclusion, as long as the coalescence is reached in a lapse that is not significantly larger than a few hundreds of Myr, the search is not affected.
Nonetheless, for completeness, an improvement of this work could take into account possible longer intervals of time, both for the selection of B-galaxies and for the assignation of B- and PTA-probabilities.

The searches described in Section \ref{sec:searches} can be improved in several ways.
One of the main shortcomings is in the method used to construct the adapted catalog.
We are assuming that all galaxies with the same redshift and mass are affected by the same observational limitations.
Moreover, we assume that the same limitations hold for galaxies that recently suffered a major merger and galaxies that did not.
If we consider that the latter assumption is not too crude, then our adaptation method should be good enough when calculating the efficiency of the simple Bayesian search explained in Section \ref{subsec:bayesian_search}.
This is so because the efficiency of the search depends only on the distribution of galaxies in $z$ and $m$, which we know for our real catalog, regardless of the observational processes that caused the distribution to be so.
When considering the galaxy clustering as a piece of information for the search, the adaptation method becomes crucial.
Nevertheless, the NNDS does not ameliorate the search performed using only $z$ and $m$ as parameters.
Therefore, even if the clustering information is affected by our adaptation method, the inclusion of the NNDS in a search on a properly adapted catalog is not expected to be efficient.
There may be yet other ways to improve the searches by including the clustering; as shown in Sections \ref{subsec:nnds} and \ref{subsec:search_clustering} there are indeed some features contained in the number of neighbors that distinguish B- from N-galaxies.

Another drawback of this work is the incompleteness of our real catalog.
The spectroscopic SDSS catalog covers a small fraction of the sky.
It would be interesting to apply the algorithms of this paper to a full sky survey.
With such a catalog we could do a more complete study of the distribution of PTA-galaxy candidates.
Data from Pan-STARRS \citep{TonryEtAl2012,SchlaflyEtAl2012} will soon be publicly released.
Their coverage is of roughly three quarters of the sky, although the redshifts will be photometric (instead of spectroscopic), inferred from the four different wavebands measured, which would be a source of uncertainty in the calculations of redshifts and masses.
Additionally, the incompleteness of our real catalog at very low redshifts ($z_\text{min}<0.01$) could be easily solved by combining this with other complete catalogs of nearby galaxies.
Nonetheless, at such low redshifts we do not expect to find more than $\sim 1$ PTA-galaxy candidates.

The cosmological parameters assumed by the MS are based on WMAP 1 data, which significantly differ from more modern measurements \citep{KomatsuEtAl2011,PlanckCollaboration2013}.
According to \cite{GuoEtAl2013}, updating the simulation to the cosmological parameters of WMAP 7 does not affect galaxy clustering significantly, since the changes in the values of $\Omega_m$ and $\sigma_8$ (the amplitude of mass fluctuations at 8$h_0^{-1}$\,Mpc) effectively compensate each other, at least below $z \lesssim 3$.
Overall, they report a small difference in the outcomes of the simulation for the two sets of cosmological parameters.
Nevertheless, an update of this work to the most recent models would be a sensible follow up.
Also, to avoid our method to be too dependent on the particular realization of the universe provided by the MS, we could redo our calculations using other simulations, like DEUS\footnote{\url{http://www.deus-consortium.org/}} \citep{AlimiEtAl2012}.

\begin{figure}
\includegraphics{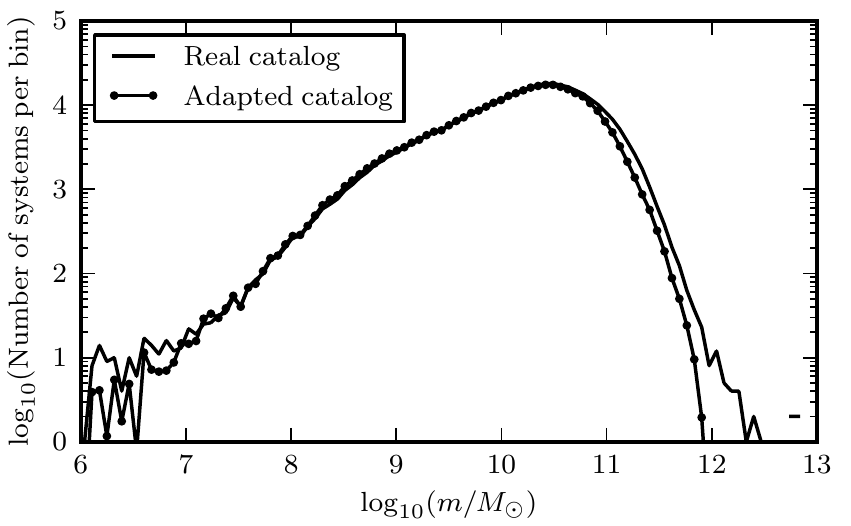}
\caption{Mass histograms of systems in the real and adapted catalogs (taking a fraction of the sky in the adapted catalog that is equal to the area covered by the real one).
Here we see that the MS has an underdensity of high-mass galaxies with respect to the SDSS.}
\label{fig:masscomp}
\end{figure}

In different sections of the paper, we have mentioned that the adapted catalog contains fewer high-mass systems than the real catalog.
This difference in masses is the reason why the number of B- and PTA-galaxy candidates found in the real and adapted catalogs disagree (by less than a factor of 2).
For the calculations of stellar masses of real galaxies, it is necessary to calculate the distance to the galaxy, which depends on the cosmological parameters; therefore, the reason for the discrepancy at the high-mass end could be related to the different parameters assumed by the two catalogs.
In Figure \ref{fig:masscomp} one can clearly see that discrepancy in masses.
Again, using a simulated universe updated to the most recent cosmological parameters could be the solution for this issue.

\begin{figure}
\includegraphics{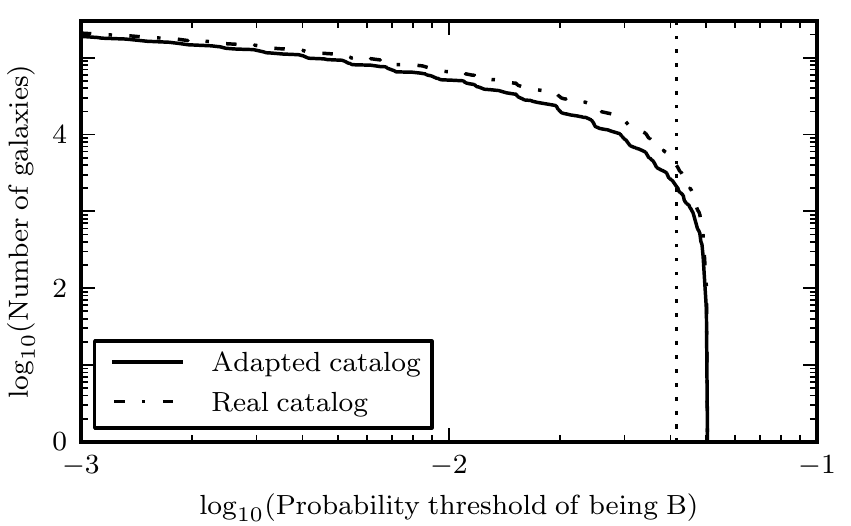}
\includegraphics{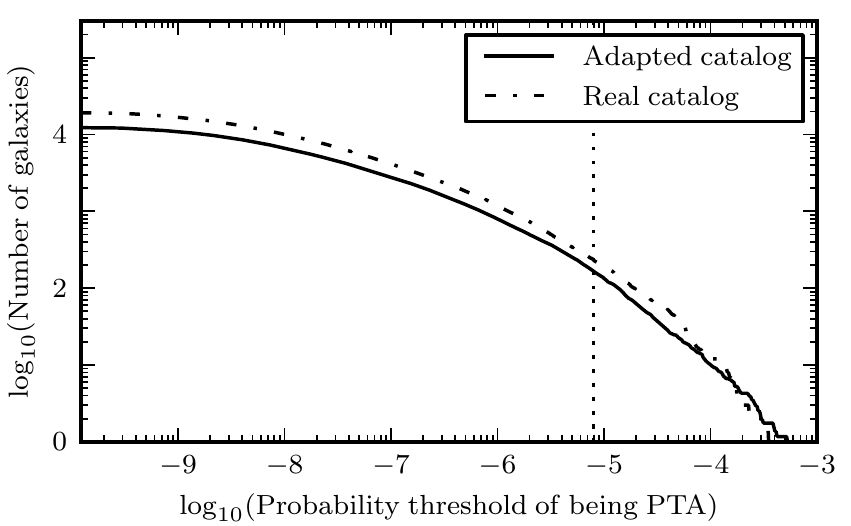}
\caption{Distribution of B-galaxy probabilities (upper plot) and PTA-galaxy probabilities (below) for the real (dot-dashed lines) and adapted (solid lines) catalogs with $z<0.1$.
The vertical axis gives the number of galaxies that have probabilities larger than a certain threshold, given on the horizontal axis.
The numbers for the adapted catalog have been multiplied by the fraction of the sky covered by the real catalog ($19.5\%$), so that solid and dot-dashed lines are comparable.
The probability thresholds $p_T^\text{B}$ and $p_T^\text{PTA}$ chosen to select B-galaxy candidates and PTA-galaxy candidates are marked with vertical dot lines.
Systems on the right of the vertical lines are B-galaxy (PTA-galaxy) candidates, in the upper (lower) plot.}
\label{fig:probhist}
\end{figure}

In Figure \ref{fig:probhist} we show the distributions of B-galaxy and PTA-galaxy probabilities (upper and lower plots, respectively) from the adapted and real catalogs.
Each point in a curve tells the number of galaxies (on the vertical axis) that have probabilities larger than a certain value (on the horizontal-axis).
The vertical lines mark the probability thresholds $p_T^\text{B}$ and $p_T^\text{PTA}$ chosen to select B- and PTA-galaxy candidates.
This plot demonstrates that probabilities are assigned in the real catalog in a similar way than in the adapted catalog, except for a factor of $\lesssim 2$ due to the mass discrepancy that we just mentioned.

\section{Conclusions}
\label{sec:conclusions}
MBHBs are expected to form in the center of massive galaxies following merger events, and the detection of their emitted GWs is the main goal of ongoing PTAs.
Whether or not MBHBs will be observed as an unresolvable background or as the sum of only a few bright signals is so far unknown, and depends on the spatial distribution of very massive systems in the low redshift universe.
As such, an effective method for predicting the properties of the PTA signal is to study the distribution of putative MBHB hosts in large galaxy surveys.

In this paper we have first investigated possible criteria to assign to each system a probability of hosting a MBHB (technically, of being a B-galaxy, i.e., a galaxy that suffered a major merger less than $\sim300\,\myr$ before emitting the radiation we observe now).
We have used a fake (simulated) galaxy catalog (result of the MS with the semi-analytical galaxy formation models of \cite{GuoEtAl2011b}) and used the peculiar two-dimensional mass-redshift distribution of merging galaxies as a selection criterion.
The fake catalog was then adapted to the observational constraints of the real catalog, the SDSS spectroscopic catalog, and the same search for B-galaxies was performed.
We caution that this method of adaptation may not be optimal (as commented on in Sections \ref{subsec:adapted_catalog} and \ref{sec:discussion}).

For each galaxy, we also calculated the probability of being a PTA-galaxy, i.e. a B-galaxy that contains a MBHB emitting GWs that produce a strain amplitude $h_0\ge h_0^\text{thres}=10^{-15}$ (in some frequency interval within the PTA frequency band).
To do this, we needed to populate galaxies with MBHs, which requires a knowledge of their bulge mass.
To infer the latter, we have constructed a simple model based on the morphological classification of the Galaxy Zoo (explained in Section \ref{subsec:real_galaxies}).

Our fiducial search is based only on the mass-redshift distribution of galaxies, and extends in the redshift interval $0.01<z<0.1$.
The search has been extended up to $z=0.7$, even though the severe incompleteness of the SDSS spectroscopic catalog at those redshifts and our simple fake catalog adaptation technique make the results of this search less robust.
We also included clustering information in our search (using a machine learning algorithm), which did not improve the efficiency of the searches for B- or PTA- galaxies. 

The main results are summarized below:
\begin{itemize}
\item In the fake universe, B-galaxies show a distinct distribution in redshift and mass (as shown in Figure \ref{fig:zmhist}): they tend to have larger masses than average (N-) galaxies, which is a reasonable consequence of the conditions in which major mergers take place. 
\item By using only this information we were able to construct a list of 3870 candidates for B-galaxies with $z<0.1$ in the SDSS footprint, of which $\sim196$ are expected to be actual B-galaxies (a skymap with these real candidates is shown in Figure \ref{fig:skymap_bgal}).
All of these have stellar masses larger than $10^{11}\,\msun$.
\item Applying our PTA-galaxy search to the real catalog, we created a list of 232 real PTA-galaxy candidates; this list is expected to include the $\sim14$ most likely PTA sources in the local universe that are observed by the SDSS (a skymap with these candidates is given in Figure \ref{fig:skymap_pta}).
PTA-galaxies also have masses $\ge 10^{11}\,\msun$.
\item According to Figure \ref{fig:tpcf_mass}, galaxies with such large masses ($\ge 10^{11}\,\msun$) are expected to have more neighbors than average galaxies, which suggests that B- and PTA-galaxies are more likely to be found in large groups or clusters.
\item The probability of actually observing these PTA-galaxy candidates is small (since they spend a relatively short interval of time producing a strain amplitude larger than $h_0^\text{thres}$), ranging from $10^{-3}$ to $10^{-5}$; nevertheless, if the PTA manages to detect single sources from the part of the sky covered by the SDSS with $z<0.1$, those sources are expected to be among the list of candidates shown in Figure \ref{fig:skymap_pta}.
\item Sensitivities to strain amplitudes smaller than $\sim 10^{-16}$ are required to have a sizable number of detectable sources.
This result supports the idea that the first PTA detection will most likely involve a low frequency stochastic background signal \citep{SiemensEtAl2013}, as opposed to a loud individual source.
\item Among the $\approx3.4\times 10^5$ galaxies in the real catalog with $z<0.1$ we are able to select $\approx1.7\times 10^4$ candidates\footnote{This number correspond to the right-most point in the black curve of Figure \ref{fig:threshold_pta}, multiplied by the factor 1.41 (explained in Section \ref{subsec:real_galaxies}) that accounts for the overabundance of SDSS massive galaxies with respect to the MS.} that should include \textit{all} PTA-galaxies ($\approx7.4\times 10^2$), if the adapted catalog resembles well the real one.
In other words, we are able to correctly classify $\approx 95\%$ of the whole galaxy population as non-PTA-galaxies, and $100\%$ of the PTA-galaxies are counted among the candidates.
Despite this great feature of the search, the number of systems misclassified as PTA-galaxies is still large (we have $\sim20$ misclassifications per true PTA-galaxy).
This is an important caveat that can be overcome (and we plan to do it in a future work) by combining our search criteria and methodology with others from the literature.
\item The search seems to keep (and even ameliorate) its efficiency when applied to redshifts larger than $\sim 0.1$, as shown in Figure \ref{fig:threshold_larger_z}.
\end{itemize}

As shown in Figure \ref{fig:numsys}, the expected number of observable PTA-galaxies in the local universe contained in our real catalog is, for a threshold of $h_0^\text{thres}=10^{-15}$, smaller than 0.01.
This number does not contradict the expected number of PTA-galaxy candidates contained in the skymap of Figure \ref{fig:skymap_pta}, which is $\sim 14$.
Among the systems in the skymap, we do expect $\sim14$ of them to be producing a strain amplitude $\ge h_0^\text{thres}$, but only during a relatively short interval of time; the sum of the PTA-galaxy probabilities of the candidates in the skymap is therefore less than 0.01.
More encouraging numbers can be achieved by setting a smaller strain amplitude threshold (to which future PTA campaigns will be sensitive), or by considering a more complete (or deeper) set of galaxies.
The upper plot in Figure \ref{fig:numsys_larger_z} is analogous to the upper plot in Figure \ref{fig:numsys}, but for redshifts up to 0.7.
The lower plot in this figure gives insight on the average number of systems that could be observed simultaneously by the PTA and an ideal telescope, able to produce a complete all-sky galaxy catalog up to 0.7.

Our work has important practical implications for MBHB and PTA-source searches.
If our understanding of the galaxy formation process is correct, there {\it must} be hundreds of binaries at $z<0.1$, and our method provides a useful way to narrow down the number of selected targets for deep imaging and spectroscopy to unveil MBHBs in the local universe.
Note, moreover, that these are massive, very low redshift systems, where the search for kinematic signatures of massive binaries at several parsec separations might be already possible with the Hubble Space Telescope\footnote{\url{www.hubblesite.org}}, and will definitely be within the capabilities of the European Extremely Large Telescope \citep{GilmozziSpyromilio2008}.
Several tens of such binaries are PTA sources, but unfortunately only a few of them will produce a strain amplitude $\ge 10^{-16}$ which might be detectable with the SKA \citep{Lazio2009}.
However, even if very small, there is a chance that a nearby galaxy hosts a loud source of GWs detectable in the PTA band, and our method proved effective in selecting the most likely candidates.
Our list of PTA-galaxy candidates can be used to perform targeted searches, also in combination with other searching criteria \citep{EllisonEtAl2013b}, for example, by looking for signs of recent galaxy interaction in the SDSS images.
Finally, being able to assign a probability to each galaxy in the universe is a powerful tool for constructing PTA signals with the `right' spatial properties.

The SDSS spectroscopic catalog covers $\sim 20\%$ of the sky, but surveys like Pan-STARRS and LSST \citep{IvezicEtAl2011} will cover almost the entire sky.
Our technique applied to all-sky, deep galaxy catalogs will allow a complete study of the expected properties (in terms of number and location of putative resolvable sources and level of anisotropy) of the low frequency GW signal in the PTA band.
This has a double value for the PTA community: on the one hand, it will provide useful guidance to the development of data analysis algorithms to search for GWs in PTA data; on the other hand, in the presence of a detection, it will provide a useful tool to interpret the results from an astrophysical standpoint.

\section*{Acknowledgments}
We thank Roberto Decarli, Iraklis Konstantopoulos and Jarle Brinchmann for the comments and suggestions; we also thank the latter for his guidance on the use of the real catalog.
P. A. R. wants to thank Bruce Allen, Colin Clark, Rutger van Haasteren, David Keitel, Drew Keppel, Reinhard Prix, Francesco Salemi, Miroslav Shaltev, and especially Tito Dal Canton for the fruitful discussions and help while the preparation of this work.
P. A. R. also thanks Guinevere Kauffmann for the valuable advices and the bibliography suggested.

The Millennium Simulation databases used in this paper and the web application providing online access to them were constructed as part of the activities of the German Astrophysical Virtual Observatory (GAVO).

Funding for the SDSS and SDSS-II has been provided by the Alfred P. Sloan Foundation, the Participating Institutions, the National Science Foundation, the U.S. Department of Energy, the National Aeronautics and Space Administration, the Japanese Monbukagakusho, the Max Planck Society, and the Higher Education Funding Council for England.
The SDSS Web Site is \url{http://www.sdss.org/}.

The SDSS is managed by the Astrophysical Research Consortium for the Participating Institutions.
The Participating Institutions are the American Museum of Natural History, Astrophysical Institute Potsdam, University of Basel, University of Cambridge, Case Western Reserve University, University of Chicago, Drexel University, Fermilab, the Institute for Advanced Study, the Japan Participation Group, Johns Hopkins University, the Joint Institute for Nuclear Astrophysics, the Kavli Institute for Particle Astrophysics and Cosmology, the Korean Scientist Group, the Chinese Academy of Sciences (LAMOST), Los Alamos National Laboratory, the Max-Planck-Institute for Astronomy (MPIA), the Max-Planck-Institute for Astrophysics (MPA), New Mexico State University, Ohio State University, University of Pittsburgh, University of Portsmouth, Princeton University, the United States Naval Observatory, and the University of Washington.

This work was supported by the IMPRS on Gravitational Wave Astronomy.

\appendix

\section{Queries for the data}
\label{sec:queries}
The SQL query sent to the MS site to download the fake catalog (for systems with redshifts $z<0.11$) is the following:
\begin{verbatim}
select h.galID,
       h.ra,
       h.dec,
       h.z_geo,
       h.z_app,
       g.stellarmass,
       g.blackholemass,
       g.bulgemass,
       g.type

from Henriques2012a.wmap1.bc03_AllSky_001 h, 
     Guo2010a..MR g

where h.galID = g.galaxyid
  and g.stellarmass > 7e-5
  and h.z_geo <= 0.11
\end{verbatim}
The stellar mass in the database is given in units of $10^{10}\,\msun/h_{100}$, where $h_{100}=0.73$ (for the set of cosmological parameters assumed by the MS), so the minimum mass imposed in this query is of $\approx9.6\times10^5\,\msun$.
Once the catalog is downloaded, we select only galaxies with masses $\ge 10^6\,\msun$ (minimum mass of the real catalog).

This query is limited to redshifts smaller than 0.11, and outputs $\approx 1.0 \times 10^7$ galaxies.
A query with a maximum redshift of 0.7 would produce $\approx 1.8 \times 10^8$ galaxies.
As we explain in Section \ref{subsec:extending}, we do not need to download all these data to extend the search to all redshifts considered in the real catalog.
Instead, we just need a histogram with the number of galaxies contained in each $z$-$m$-bin.

The searches described in Sections \ref{subsec:bayesian_search} and \ref{subsec:search_pta} are restricted to $z<0.1$.
However, we download systems with $z<0.11$ to avoid border effects: when calculating the NNDS of systems close to $z=0.1$, one needs to consider background galaxies that have larger redshifts; for our choices of shells, all background galaxies are safely contained below $z=0.11$.

The query used to construct the list of B-galaxies in the fake catalog is:
\begin{verbatim}
select h.galID,
       h.z_geo,
       h.z_app,
       p1.blackholemass,
       p2.blackholemass

from Henriques2012a.wmap1.bc03_AllSky_001 h,
     Guo2010a..MR p1,
     Guo2010a..MR p2,
     Guo2010a..MR d 

where h.galID = d.galaxyid
  and h.z_geo <= 0.7
  and p1.descendantid = d.galaxyid
  and p2.descendantid = d.galaxyid
  and p1.galaxyid < p2.galaxyid
  and p1.snapnum = p2.snapnum
  and p1.stellarmass >= 0.2*d.stellarmass
  and p2.stellarmass >= 0.2*d.stellarmass
  and p1.blackholemass > 1e-6
  and p2.blackholemass > 1e-6
  and d.blackholemass > 1e-6
  and d.stellarmass > 7e-5
  and p1.disruptionon = 0
  and p2.disruptionon = 0
  and p1.snapnum = d.snapnum-1
\end{verbatim}
This query outputs the \verb|galID| of galaxies that suffered a major merger between their snapshot and the previous one.
Since the \verb|galID| can be repeated (as mentioned in Section \ref{subsec:fake_catalog}), cosmological and apparent redshifts are also downloaded, to be able to identify systems without any ambiguity.
The progenitors' black hole masses are downloaded to construct the black hole mass ratio, necessary for the calculations presented in Section \ref{subsec:search_pta}.

The condition of major merger is that at least two progenitors of a galaxy must have mass $\ge0.2$ times the mass of the galaxy.
Furthermore, we discard progenitors that were disrupted before the merger.
We also impose that the progenitor must have a black hole mass larger than $\approx 1.4\times 10^4\,\msun$.
The same condition is imposed to the descendant, although this condition turns out to be unnecessary, since the minimum black hole mass found among B-galaxies is of $\approx 10^{6.2}\,\msun$.

Several major multimergers are found with this query.
These are mergers of three galaxies\footnote{The unlikely cases of major multimergers involving more than three progenitors were not found with this query.} that have a mass larger than 0.2 times the mass of the descendant.
When such a multimerger occurs, the query outputs the descendant galaxy three times, because three possible pairs of progenitors are considered (first and second, first and third, and second and third).
If that descendant galaxy appears twice (or three times, or four times, etc.) in the galaxy catalog (due to a repetition of the cube of the simulation), the query will output 6 (or 9, or 12, etc.) times the same galaxy.
One has to properly correct for these repetitions to avoid ambiguities when identifying galaxies.

The query used in the SDSS DR7 server to obtain the morphological parameters introduced in Section \ref{subsec:real_galaxies} is the following:
\begin{verbatim}
select g.objid,
       s.specobjid,
       g.petror50_r,
       g.petror90_r,
       g.petror50_z

from galaxy g,
     specobj s

where g.objid=s.bestobjid
  and s.z<=0.7
\end{verbatim}
The two first items, \verb|objid| and \verb|specobjid| were used to identify the galaxies of this query with the ones of our real catalog.

\bibliographystyle{mn2e}
\bibliography{observations}

\label{lastpage}

\end{document}